\documentclass[10pt,letterpaper,twocolumn]{article}
\usepackage[english]{babel}
\usepackage{graphicx}

\newcommand{\alt}{\mathbin{\lower 3pt\hbox
   {$\rlap{\raise 5pt\hbox{$\char'074$}}\mathchar"7218$}}}
\newcommand{\agt}{\mathbin{\lower 3pt\hbox
   {$\rlap{\raise 5pt\hbox{$\char'076$}}\mathchar"7218$}}}

\begin{document}

\setcounter{footnote}{0}
\setcounter{equation}{0}
\setcounter{figure}{0}
\setcounter{table}{0}

\title{\large\bf Boundary conditions, phase distribution \\
and hidden symmetry in 1D localization }

\author{\small I. M. Suslov  \\
\small P.L.Kapitza Institute for Physical Problems,  \\
\small 119334 Moscow, Russia  \\
\small E-mail: suslov@kapitza.ras.ru\\
 {}\\
\parbox{150mm}{\footnotesize \,One-dimensional disordered
systems with a random potential of a small amplitude and
short-range correlations are considered near the initial
band edge.  The evolution equation is obtained for the mutual
ditribution $P(\rho,\psi)$ of the Landauer resistance $\rho$
and the phase variable $\psi=\theta-\varphi$ ($\theta$ and
$\varphi$ are phases entering the transfer matrix), when the
system length $L$ is increased. In large $L$ limit, the
equation allows separation of variables, which provides the
existence of the stationary distribution $P(\psi)$,
determinative the coefficients in the evolution equation for
$P(\rho)$. The limiting distribution $P(\rho)$ for $L\to\infty$
is log-normal and does not depend on boundary
conditions. It is determined by the 'internal' phase
distribution, whose form is established in the whole energy
range including the forbidden band of the initial crystal.
The random phase approximation is valid in the deep of
the allowed band, but strongly violated for other energies.
The phase $\psi$ appears to be the 'bad' variable, while the
'correct' vaiable is $\omega=-{\rm ctg}\,\psi/2$. The form
of the stationary distribution $P(\omega)$ is
determined by the internal properties of the system and is
independent of boundary conditions. Variation of the boundary
conditions leads to the scale transformation $\omega\to s\omega$
and translations $\omega \to \omega+\omega_0$ and
$\psi\to\psi+\psi_0$, which determinates the 'external' phase
distribution, entering the evolution equations. Independence of
the limiting distribution $P(\rho)$ on the external
distribution $P(\psi)$ allows to say on the hidden
symmetry, whose character is revealed below.  } }

\date{}
\maketitle


\setcounter{footnote}{0}
\setcounter{equation}{0}
\setcounter{figure}{0}
\setcounter{table}{0}

\begin{center}
{\bf 1. Introduction}
\end{center}

For description of 1D disordered systems
it is convenient to use the transfer matrix $T$, relating
the amplitudes of plane waves on the left ($Ae^{ikx}+Be^{-ikx}$)
and on the right ($Ce^{ikx}+De^{-ikx}$) of a scatterer,
$$
\left ( \begin{array}{cc} A \\ B \end{array} \right)\,
=  T \left ( \begin{array}{cc} C \\ D \end{array}\right)
\,.
\eqno(1)
$$
In the presence time-reversal invariance, the matrix $T$
can be parametrized in the form \cite{1}
$$
 T= \left ( \begin{array}{cc} \!\!\! 1/t\! &\! - r/t \!\!\\
\!\!- r^*/t^* \!&\! 1/t^* \!\!\!\end{array} \right)\,
= \left ( \begin{array}{cc}
\!\!\sqrt{\rho\!+\!1}\, e^{i\varphi}\!\! &
\!\!\sqrt{\rho} \,e^{i\theta}\!\!
\\ \!\!\sqrt{\rho}\, e^{-i\theta}\!\!
&\!\! \sqrt{\rho\!+\!1}\,
e^{-i\varphi}\!\! \end{array} \right)\,,
\eqno(2)
$$
where $t$ and $r$ are the amplitudes of transmission
and reflection, while $\rho=|r/t|^2$ is the dimensionless
Landauer resistance \cite{2}.
 For the successive
arrangement of scatterers their transfer matrices are
multiplied. For a weak scatterer its transfer matrix $T$ is
close to the unit one, which allows to derive the differential
evolution equations for its parameters, and in particular
for the Landauer resistance $\rho$.

In the random phase approximation (when distributions of
$\varphi$ and  $\theta$ are considered as uniform) such
equation for the distribution $P(\rho)$ has a form
\cite{3}--\cite{8}
$$
\frac{\partial P(\rho)}{\partial L} =
D\,\frac{\partial}{\partial \rho}
\left[\,\rho(1\!+\!\rho)\,\frac{\partial P(\rho)}{\partial \rho}
\,\right] \, \eqno(3)
$$
(where $D$ is of the order of the inverse mean free path)
and describes evolution of the initial distribution
$P_0(\rho)=\delta(\rho)$ at zero length $L$ to the log-normal
distribution in the large $L$ limit.

As shown in the paper \cite{9},
the distributions of phases $\varphi$
and $\theta$ ceased be uniform, if
semi-transparent boundaries are introduced between the
disordered system and the ideal leads connected to it, even
if they were uniform in the initial system. In the latter case,
the more general equation arises
$$
\frac{\partial P(\rho)}{\partial L} =
D\,\frac{\partial}{\partial \rho}
\left[\,-\gamma(1\!+\!2\rho) P(\rho) +
\rho(1\!+\!\rho)\,\frac{\partial P(\rho)}{\partial \rho}
\,\right]   \,,
\eqno(4)
$$
which reduces to (3) in the random phase approximation.
The latter approximation is working sufficiently good in
the deep of the allowed band for the "natural" ideal
leads (made from the same material as a disordered
system, but without impurities),
as it is usually accepted in the theoretical papers (see
references in \cite{10,10a,11});
the fluctuation states in the
forbidden band are considered infrequently \cite{12,13,14} and
only on the level of the wave functions.  To study the evolution
of $P(\rho)$ for the arbitrary Fermi level position  (including the
forbidden band of the initial crystal), one should explicitly
introduce the foreign ideal leads made from the good
metal\footnote{\,It is easy to understand, that for energies in
the forbidden band of the initial crystal the 'natural' ideal leads
will be non-conducting, while the fluctuational states are present
in the disordered system under consideration.},
which automatically brings to the nontrivial boundary conditions.
As a result, the still more general equation arises \cite{15},
$$
\frac{\partial P(\rho)}{\partial L} =
D\,\frac{\partial}{\partial \rho}
\left[\,-\gamma_1(1\!+\!2\rho) P(\rho)-
\vphantom{\frac{1}{2}} \right.
$$
$$  \left.
\,-2\gamma_2 \sqrt{\rho(1\!+\!\rho)} P(\rho)
+\rho(1\!+\!\rho)\,\frac{\partial P(\rho)}{\partial \rho}
\,\right]   \,,
\eqno(5)
$$
whose coefficients are determined by the stationary phase
distribution (see Eq.\,34 below) in the large $L$ limit.
Equation (5) reduces to (4) with $\gamma=\gamma_1+\gamma_2$
in the region of large $L$, when typical values of $\rho$ are
large. Fig.1 illustrates the dependence of the
parameter $\gamma$ on the quantity
$\tilde{\cal E}={\cal E}/W^{4/3}$, where
${\cal E}$ is the Fermi energy counted from the lower edge of the
initial band, and $W$ is the amplitude of a random potetial;
all energies are measured in the units of the hopping integral
for the 1D Anderson model (see below Eq.13).
Parameter $\gamma$ is close to zero in the deep of the allowed
band in correspondence with the random phase approximation,
while violation of the latter in the whole energy interval is
surely not a small effect. This violation occurs due to internal
reasons and incorporation of semi-transparent boundaries is
not necessary for it.
Moreover, the  limiting log-normal distribution $P(\rho)$ in the large
$L$ region,
$$
P(\rho)=\frac{1}{\rho \sqrt{4\pi D L}}
\exp\left\{-\frac{[\ln \rho-vL]^2}{4DL}\right\}\,,
\eqno(6)
$$
where $v=(2\gamma\!+\! 1)D$, is determined by the internal properties
of the disordered system (Fig.2) and does not depend on the boundary
conditions \cite{15}.

\begin{figure}
\centerline{\includegraphics[width=3.4 in]{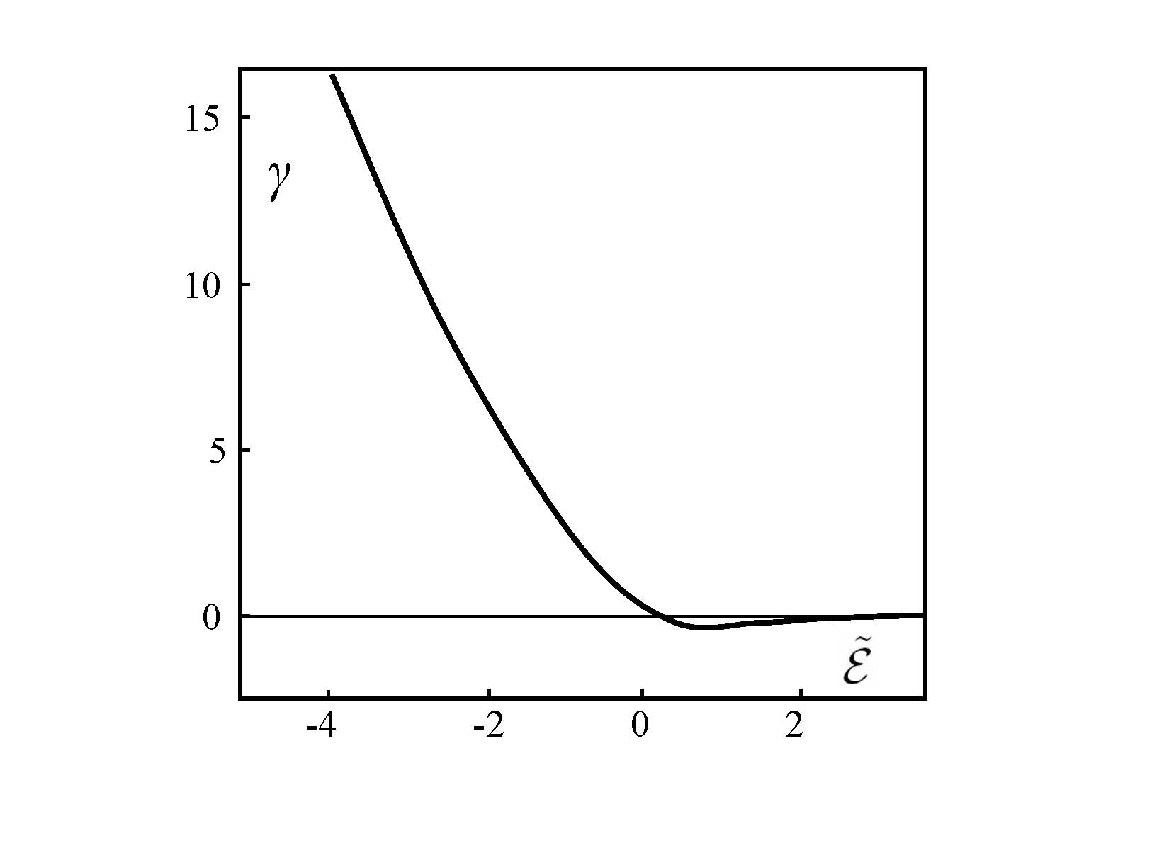}}
\caption{ Parameter $\gamma$ in equation (4), as a function
 of the reduced energy $\tilde{\cal E}={\cal E}/W^{4/3}$.
Parameter $\gamma$ is close to zero in the deep of the allowed
band, in correspondence with the random phase approximation.
} \label{fig1}
\end{figure}

The dependence of $\gamma$ on  $\tilde{\cal E}$ (Fig.1) clearly
demonstrates violation of the random phase approximation, but
was obtained in \cite{15} by analysis of the moments of the
transfer matrix elements (see below), in which the
problem of phase distribution was completely avoided.
Clarification of a situation with  phase distribution is
necessary for the logical completeness of theory of 1D
localization and is the main purpose of the present paper. It is
closely related with elucidation of the role of boundary
conditions and needs resolution of  visible contradictions.

\begin{figure}
\centerline{\includegraphics[width=3.4 in]{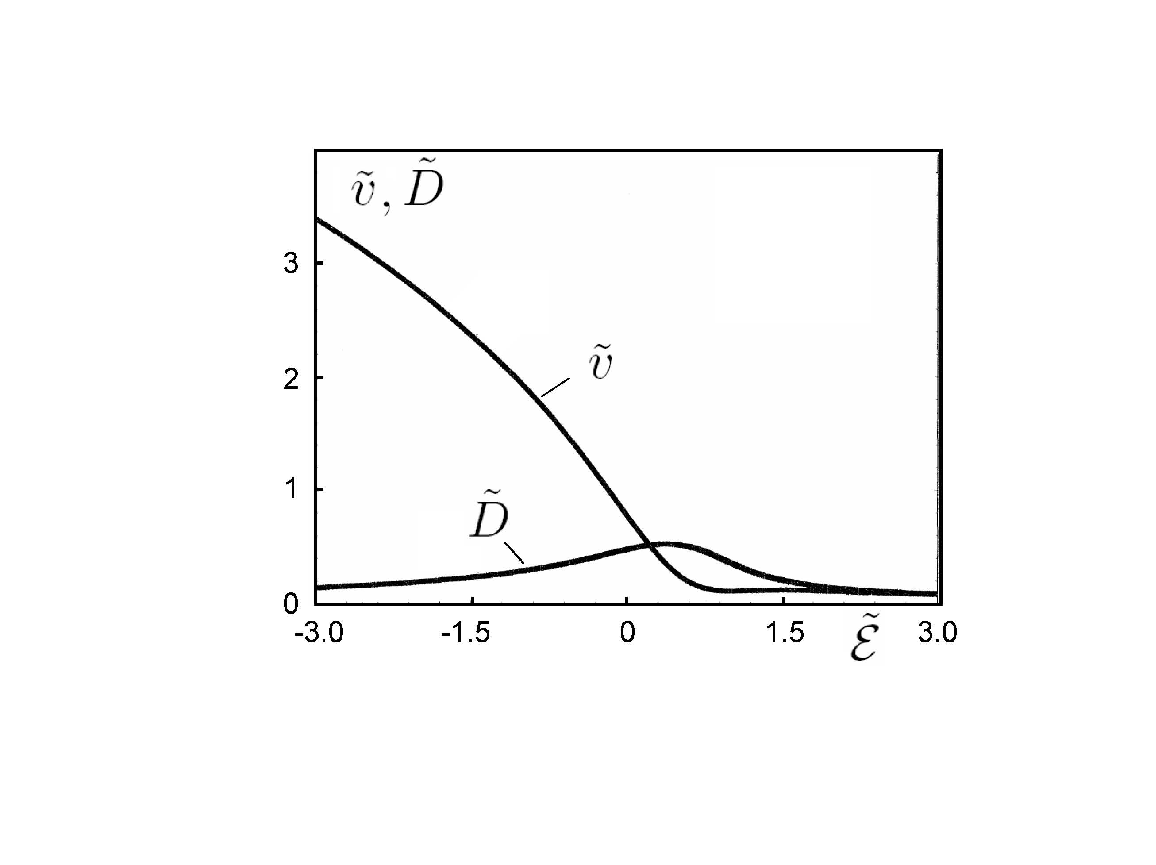}} \caption{
Parameters $\tilde v=v/W^{2/3}$ and $\tilde
D=D/W^{2/3}$ versus the reduced energy $\tilde{\cal E}={\cal E}/W^{4/3}$.
The equality $v=D$, followed from the random phase approximation, is realized
only in the deep of the allowed band.
} \label{fig2}
\end{figure}

Inddeed, in the papers \cite{9,15} we have made two statements,
which look hardly compatible. On one hand,
variation of the boundary conditions essentially affects
the distribution of phases, which generally
changes the parameters
of the evolution equations (3--5) and even its structure.
On the other hand, these changes have no influence on the form of
the limiting distribution (6) in the large $L$ region.
Validity of these two statements means that the system obeys
a hidden symmetry, i.e. invariance of the physical quantities
respective to a certain class of transformations.
From the
theoretical viewpoint, revelation of the hidden symmetry
is of the evident interest, indicating the possibility of
essential simplifications. From the practical point,
one cannot differ the real physical effects from the fictive
ones, if the nature of hidden invariance is not clarified.
Revelation of this invariance appears to be very nontrivial
and demands derivation of the evolution equations in the
most general form.

Let explain the origin of two indicated statements. Under
a change of the boundary conditions, the transfer matrix $T$
transforms to $\tilde T=T_l T T_r$, where $T_l$ and $T_r$
are the edge matrices, related amplitudes of waves on the left
and on the right of the corresponding interface. Thereby, the
change of the boundary conditions leads to the linear
transformation of the transfer matrix elements.
The linear
transformation does not affect the growth exponents
for the second and forth moments of the matrix
elements, which can be found for a given matrix $T$ and
hence are
determined by internal properties of the system. Knowledge
of these two exponents allows to establish the 'diffusion
constant' $D$ and  the 'drift velocity' $v$ in the limiting
distribution (6) (Fig.2), which consequently does not depend on the
boundary conditions \cite{15}; after that there is no problem
to obtain the behavior of the parameter $\gamma$ (Fig.1).

Influence of boundary conditions on the  distribution of phases
can be easily demonstrated by introducing the point scatterers on
the system boundaries, when
$$
\tilde T=T_l T T_r\,,\qquad
T_l=T_r=
\left ( \begin{array}{cc} 1\!-\!i\chi & -i\chi \\
         i\chi & 1\!+\!i\chi \end{array} \right)\,.
\eqno(7)
$$
Accepting the parametrization (1) for $\tilde T$,
one has in the main order for large $\chi$
$$
\sqrt{1\!+\!\rho} \,{\rm e}^{i\varphi} = -\chi^2\,  T'\,, \qquad
\sqrt{\rho}\, {\rm e}^{i\theta} = -\chi^2\,  T'\,,
$$
$$
\sqrt{\rho}\, {\rm e}^{-i\theta} = \chi^2 \, T'\,, \qquad
\sqrt{1\!+\!\rho}\, {\rm e}^{-i\varphi} = \chi^2 \, T'\,,
\eqno(8)
$$
where $T'=T_{11}\!-\!T_{12}\!+\!T_{21}\!-\!T_{22}$ and $T_{ij}$
are the elements of the $T$-matrix.
For large $\chi$ we have $\rho\sim \chi^4$ and
$1\!+\!\rho \approx \rho$, so it is easy to see
that
$$
\varphi =\pm\pi/2, \quad \theta =\pm\pi/2  \qquad
\mbox{\rm for}\,\quad \chi \to \infty\,.
\eqno(9)
$$
Thereby, for large $\chi$  the phase variables $\varphi$ and
$\theta$ are localized near values $\pm \pi/2$ independently
of their distributions in the initial system.

\begin{figure*}
\centerline{\includegraphics[width=5.0 in]{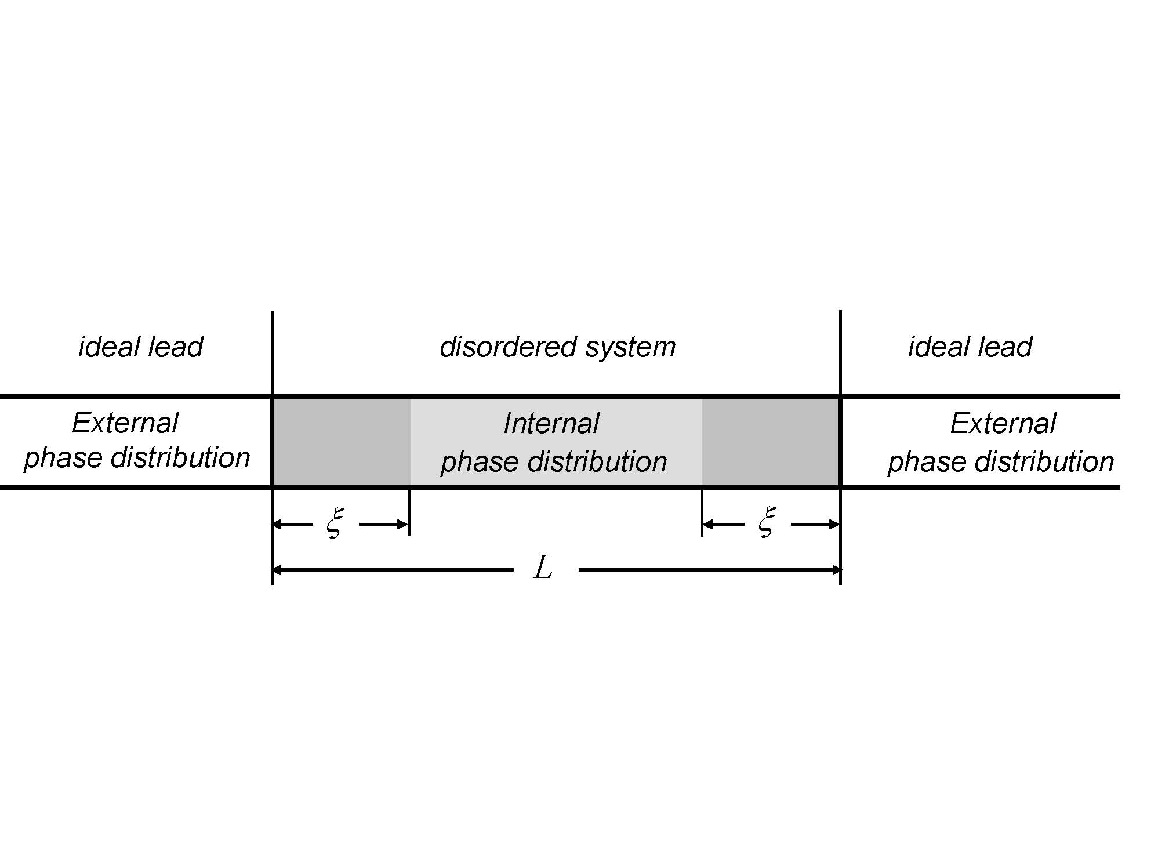}}
\caption{
External and internal phase distribution.
} \label{fig3}
\end{figure*}

From the physical viewpoint, the situation looks as follows
(Fig.3). In the deep of the sufficiently long disordered system,
a certain 'internal' phase distribution is realized, which does not
depend on the boundary conditions. If the system is considered
from the side of the ideal leads, one observes the 'external' phase
distribution, which is determined by the boundary conditions; namely
these phases are entered in the transfer matrix. Influence of interfaces
extends till the length scale of the order of the localization length
$\xi$, which determined the transient region, where the internal
phase distribution continuoualy transforms to the external one.
In the large $L$ limit, the distribution $P(\rho)$ is determined by
the internal phase distribution, which provides its independence on the
boundary conditions. However, the evolution equations contain namely
the external phase distribution\,\footnote{\,It is quite natural, since
for small $L$ the internal phase distribution is not manifested at all,
and the boundary conditions essentially affect the distribution
$P(\rho)$, as was extensively discussed in  \cite{15}.},
and there is a problem  to understand,
why it does not affect the limiting distribution $P(\rho)$. The second
question, related with the first one, is as follows: how can we
find the internal phase distribution, if it is not entering the
evolution equations?

Let discuss the character of invariance mentioned above.
The change of the matrix $T$ with a system length $L$
is determined by relation
$$
T_{L+\Delta L}=T_{L}\,T_{\Delta L} \,,
\eqno(10)
$$
where the matrix $T_{\Delta
L}$ is close to the unit one; it allows to derive the
differential evolution equations. For the change of boundary
conditions, let multiply Eq.\,10 by $T_l$ and $T_r$,
introducing the product $T_r T_r^{-1}=1$ between two multipliers:
$$
T_{l}\,T_{L+\Delta L}T_{r}=T_{l}T_{L}T_{r}\cdot
T^{-1}_{r}T_{\Delta L}T_{r} \,.
\eqno(11)
$$
Then for the matrix $\tilde T_{L}=T_{l} T_L T_{r}$ one has the relation,
analogous to (10)
$$
\tilde T_{L+\Delta L} = \tilde T_{L} \, T'_{\Delta L}\,,
\eqno(12)
$$
where
the matrix $T'_{\Delta L}=T^{-1}_{r}T_{\Delta L}T_{r}$
is again close to the unit one. A passage from
$T_{\Delta L}$ to $T'_{\Delta L}$ changes the form of the
evolution equations, determined by the parameters
$\alpha$, $\beta$, $\gamma$, $\Delta$, $\epsilon^2$
(see Secs.\,2,\,3),
while a passage from  $T_{L}$ to $\tilde
T_{L}$ changes the stationary phase distribution, which
determines the coefficients in Eq.\,5 for $P(\rho)$.
These two factors should compensate each other, in order
the limiting distribution $P(\rho)$ remains invariant.

However, such invariance is not evident from the evolution
equations, and its revelation needs essential efforts. Resolution
of these difficulties is closely related whith solution of the
question on the internal phase distribution.

\begin{center}
{\bf 2. Succession of point scatterers}
\end{center}

As clear from experience of the paper \cite{15}, it is
convenient to consider the energies incide the forbidden
band of the initial crystal, while the description of the
allowed band can be obtained by analytical continuation.
For definiteness, we have in mind the 1D Anderson model
$$
\Psi_{n+1}+\Psi_{n-1}+V_n \Psi_n = E \Psi_n \,
\eqno(13)
$$
near the band edge, where it corresponds to discretization
of the usual continuous Schroedinger equation; $E$ is the energy
counted from the band center.

A scatterer in the forbidden band is described by the
pseudo-transfer matrix $t$, relating solutions on the left
($Ae^{\kappa x}+Be^{-\kappa x}$) and on the right
($Ce^{\kappa x}+De^{-\kappa x}$) of the scatterer.
Succession of scatterers with amplitudes
$V_0$,\,$V_1$,\,$V_2$,\,$\ldots$,\,$V_n$, arranged at
the points $0$, $L_1$, $L_1\!+\!L_2$,
$\ldots\,\,$, $L_1\!+\!L_2\!+\!\ldots\!+\!L_n$, is described
by the matrix
$$
t^{(n)}= t_{\epsilon_0} \, t_{\delta_1} \, t_{\epsilon_1}\,
t_{\delta_2}\,t_{\epsilon_2}\,
\ldots\,
 t_{\delta_n}\, t_{\epsilon_n}\,,
\eqno(14)
$$
where
$$
 t_{\epsilon_n}=
\left ( \begin{array}{cc} 1+\bar\epsilon_n & \bar\epsilon_n \\
-\bar\epsilon_n & 1-\bar\epsilon_n \end{array} \right)\,,\qquad
\bar\epsilon_n =\frac{V_n}{2\kappa a_0}\,,
\eqno(15)
$$
$$
t_{\delta_n}= \left ( \begin{array}{cc}
{\rm e}^{-\delta_n} & 0 \\ 0 & {\rm e}^{\delta_n} \end{array}
\right)\,,\qquad \delta_n =\kappa L_n\,
$$
and $a_0$ is the lattice constant. The passage to the true
transfer matrix $T^{(n)}=T_l\,\, t^{(n)} \,T_r$ is realized
with the help of the edge matrices,
describing the attachment of the
ideal leads made from the good metal with the Fermi
momentum $k$. Introducing the
product $ T_r T_l=1$ between any two multipliers in Eq.14, we
have
\,\footnote{\,The condition $ T_r T_l=1$ can be accepted without
the loss of generality. If it is not so, then we can set
$ T_l= T'_l T''_l$ with $ T_r T''_l=1$, and use the matrix
$ T''_l$ instead of $ T_l$.  The role of the matrix $ T'_l$
reduces to the change of the initial conditions to the evolution
equation, while the form of the latter is not changed. }
$$
 T^{(n)}= T_{\epsilon_0}
\, T_{\delta_1} \, T_{\epsilon_1}\,
T_{\delta_2}\,T_{\epsilon_2}\,
\ldots\,
 T_{\delta_n}\, T_{\epsilon_n}\,,
\eqno(16)
$$
where
$$
T_{\epsilon_n}=  T_l \,\,t_{\epsilon_n}\,  T_r \,,\qquad
T_{\delta_n}=  T_l \,\,t_{\delta_n}\,  T_r\,.
\eqno(17)
$$
In the Anderson model all $\delta_n$ are equil,
$\delta_n=\kappa a_0$, since the scatterers are present at each
site of the lattice. Examples of the edge matrices are given
below, and in the most general case lead to the following
structure of the matrices  $T_{\epsilon_n}$ and $T_{\delta}$
$$
 T_{\epsilon_n}=
\left ( \begin{array}{cc} 1\!-\!i\epsilon_n
& \epsilon_n {\rm e}^{i\gamma} \\
\epsilon_n {\rm e}^{-i\gamma} &
1\!+\!i\epsilon_n \end{array} \right)\,,\qquad
\epsilon_n=K \bar \epsilon_n \,,
\eqno(18)
$$
$$
 T_{\delta}= \left ( \begin{array}{cc}
{\cal A} & {\cal B} \\ {\cal B}^* & {\cal A}^* \end{array} \right)\,=
\left ( \begin{array}{cc}
\sqrt{1\!+\!\Delta^2}\, {\rm e}^{i\alpha} \!\!  &
\Delta {\rm e}^{i\beta} \!\!\!
\\ \Delta {\rm e}^{-i\beta} \! & \sqrt{1\!+\!\Delta^2}\, {\rm
e}^{-i\alpha}\!\! \end{array} \right).
$$
The matrix $T_{\delta}$ is the transfer matrix of the general form,
while for $T_{\epsilon_n}$ the given form is sufficient.
As usual, we accept that all $V_n$ are
statistically independent, and $\langle
V_n \rangle=0$, $\langle V^2_n \rangle=W^2$.
Then the evolution equations will be determined by parameters
$\alpha$, $\beta$, $\gamma$, $\Delta$ and the quantity
$$
\epsilon^2=\langle \epsilon^2_n \rangle=
{\rm const}\, W^2\,.
\eqno(19)
$$
In the allowed band for the 'natural' ideal leads, the
succession of point scatterers is described not by product (14),
but the product (16), where
$$
 T_{\epsilon_n}=
\left ( \begin{array}{cc} 1-i\epsilon_n & -i\epsilon_n \\
i\epsilon_n & 1+i\epsilon_n \end{array} \right)\,,\qquad
\epsilon_n =\frac{V_n}{2\bar k a_0}\,,
$$
$$
T_{\delta_n}= \left ( \begin{array}{cc}
{\rm e}^{-i\delta_n} & 0 \\ 0 & {\rm e}^{i\delta_n} \end{array}
\right)\,,\qquad
\delta_n =\bar k L_n\,,
\eqno(20)
$$
and $\bar k$ is the Fermi momentum in our disordered system, which
corresponds to the change $\kappa \to i\bar k$ in the previous relations.
Such change corresponds to the smooth transition from the energy
${\cal E}=-\kappa^2$ in the forbidden band to the energy
${\cal E}=\bar k^2$ in the allowed band. The change of the boundary
conditions leads to the matrices $\tilde T_{\epsilon_n}=T_l T_{\epsilon_n}T_r$
and $\tilde T_{\delta_n}=T_l T_{\delta_n}T_r$, having the structure (18).

\begin{center}
{\bf 3. Evolution equations }
\end{center}

Let use the recurrence relation
$$
T^{(n+1)}=T^{(n)}T_\delta T_\epsilon \,,
\eqno(21)
$$
where matrces $T^{(n)}$ and $T_\epsilon $ are statistically
independent, and $T_\delta$ is not random.
Accepting parametrization (2) for  $T^{(n)}$, and
designating parameters of the matrix $T^{(n+1)}$
as $\tilde \rho$, $\tilde \varphi$, $\tilde \theta$,
we have
$$
\sqrt{1\!+\!\tilde\rho}\,{\rm e}^{i\tilde\varphi}
= \sqrt{1\!+\!\rho}\, {\rm e}^{i\varphi}
({\cal A}+\epsilon {\cal C}) +\sqrt{\rho}\, {\rm e}^{i\theta}
({\cal B}^*\!+\epsilon {\cal D}^*) \,,
\eqno(22)
$$
$$
\sqrt{\tilde\rho}\,{\rm e}^{i\tilde\theta}
= \sqrt{1\!+\!\rho}\, {\rm e}^{i\varphi}
({\cal B}\!+\!\epsilon {\cal D}) +\sqrt{\rho}\, {\rm e}^{i\theta}
({\cal A}^*\!+\epsilon {\cal C}^*) \,,
$$
where we introduced notations
$$
{\cal C}=\!{\cal B}\,{\rm e}^{-i\gamma}\!-\!i{\cal A}\,
\qquad
{\cal D}=\!{\cal A}\,{\rm e}^{i\gamma}\!+\!i{\cal B}\,.
\eqno(23)
$$
In what follows we consider the limit
$$
\delta\to 0\,, \quad \epsilon \to 0\,,\quad
\delta/\epsilon^2=const
\eqno(24)
$$
and retain the terms of the first order in $\delta$ and the
second order in $\epsilon$. Squaring the modulus of one of
equations (22), we have
$$
\tilde\rho= \rho+{\cal K} \sqrt{\rho(1\!+\!\rho)}
+\epsilon^2 (1\!+\!2\rho)\,, \
\eqno(25)
$$
where
$$
{\cal K}= 2\Delta\cos{(\psi\!\!-\!\!\beta)} +2\epsilon
\cos{(\psi\!\!-\!\!\gamma)} -2\epsilon^2
\sin{(\psi\!\!-\!\!\gamma)}\,
\eqno(26)
$$
and the combined phase variable is introduced
$$
\psi=\theta-\varphi\,.
\eqno(27)
$$
Now let take the product of the second equation (22)
with the complex conjugated first equation
$$
\sqrt{\tilde\rho(1\!+\!\tilde\rho) }\,{\rm e}^{i\tilde\psi}=
(1\!+\!2\rho)\,
\left[\Delta{\rm e}^{i\beta}\!+\!\epsilon{\rm e}^{i\gamma}
+\!i\epsilon^2{\rm e}^{i\gamma} \right]+
$$
$$
+ \sqrt{\rho(1\!+\!\rho)}\,
\left[\left({\rm e}^{-2i\alpha}\!
+\!2i\epsilon\!-\!\epsilon^2 \right)
\,{\rm e}^{i\psi}+
\epsilon^2\,{\rm e}^{2i\gamma-i\psi} \right].
\eqno(28)
$$
Excluding $\tilde\rho$ using equation (25), we obtain the
relation between  $\tilde\psi$ and $\psi$
$$
\tilde\psi=\psi+2\,(\epsilon\!-\!\alpha)+
(R^2/2\!-\!1)\,\epsilon^2\sin{2(\psi\!-\!\gamma)}-
$$
$$
-R\,\left[\Delta\sin{(\psi\!-\!\beta)}
+\epsilon\sin{(\psi\!-\!\gamma)}
+\epsilon^2\cos{(\psi\!-\!\gamma)} \right]\,,
\eqno(29)
$$
where
$$
R=\frac{1\!+\!2\rho}{\sqrt{\rho(1\!+\!\rho)}} \,.
\eqno(30)
$$
Using (22), (26) and following the scheme of the papers
\cite{9,15,16}, we come to the evolution equation for
$P(\rho,\psi)$
$$
\frac{\partial P}{\partial L}=
\left\{\vphantom{\frac{1}{2}}
\epsilon^2\left[1\!-\!2\cos^2{(\psi\!-\!\gamma)}\right]
(1\!+\!2\rho) P -
\right.
$$
$$
-2\left[\Delta\!\cos{(\psi\!-\!\beta)}+
\epsilon^2\sin{(\psi\!-\!\gamma)}\right]
\sqrt{\!\rho(\!1\!+\!\rho\!)} P+
$$
$$
+2\epsilon^2\! \cos\!^2{(\psi\!-\!\gamma)} \rho(\!1\!+\!\rho\!)
P'_{\rho}+
$$
$$ \left.
+2\epsilon^2\! \cos{(\psi\!-\!\gamma)}
\left[\vphantom{R^2} 2\!-\!R \sin{(\psi\!-\!\gamma)}
   \right]\sqrt{\rho(1\!+\!\rho)} P'_{\psi} \right\}'_\rho +
$$
$$
+\left\{ \vphantom{\frac{1}{2}}
\epsilon^2\cos{(\psi\!-\!\gamma)}
\left[\vphantom{R^2}  2\sin{(\psi\!-\!\gamma)}\!-\!R\right] P+
 \right.
$$
$$
+\left[\vphantom{R^2}
2\alpha\!+\!R\Delta\sin{(\psi\!-\!\beta)} \right] P +
$$
$$
\left.
+\frac{1}{2}\epsilon^2 \left[\vphantom{R^2} 2\!-
     \!R\sin{(\psi\!-\!\gamma)}\right]^2\, P'_\psi
 \right\}'_\psi \,.
\eqno(31)
$$
The right-hand side is a sum of full derivatives, which provides
the conservation of probability. Relation between distributions
of $\rho$ and $\psi$ is determined by the quantity $R$, which tends
to 2 in the limit of large $L$, when the typical values of $\rho$ are
large. Then the solution of Eq.31 is factorized, $P(\rho,\psi)=P(\rho)P(\psi)$,
though the situation is somewhat unusual for separation of variables
(see Appendix 1); the equation for $P(\psi)$ is splitted off,
$$
\frac{\partial P(\psi)}{\partial L}=
\left\{ \vphantom{\frac{1}{2}}
\left[\vphantom{R^2}
2\alpha\!+\!2\Delta\sin{(\psi\!-\!\beta)}
-2\epsilon^2\cos{(\psi\!-\!\gamma)}+\right.
\right.
$$
$$
\left.
+\epsilon^2\sin{2(\psi\!-\!\gamma)}
\right] P(\psi) +
$$
$$+
\left.
2\epsilon^2 \left[\vphantom{R^2} 1\!-
     \!\sin{(\psi\!-\!\gamma)}\right]^2\, P'_\psi(\psi)
 \right\}'_\psi \,
\eqno(32)
$$
giving the condition for the stationary distribution
of the phase $\psi$
$$
\epsilon^2 \left[\vphantom{R^2}
1\!-\!\sin{(\psi\!-\!\gamma)}\right]^2 P'_\psi-
$$
$$-
\, \epsilon^2 \left[\vphantom{R^2}
1\!-\! \sin{(\psi\!-\!\gamma)}\right]
\cos{(\psi\!-\!\gamma)} P+
$$
$$
+ \left[\vphantom{R^2} \alpha\! +\! \Delta\sin{(\psi\!-\!\beta)} \,
\right] P =C_0 \,,
\eqno(33)
$$
where the constant $C_0$ is fixed by
normalization.\,\footnote{\,Equations (31,\,32)
are analogous
to Eqs.(10.27), (10.28) in the book \cite{10}, derived in the
framework of the different formalism, so the quantities
entering them are not related clearly with the
transfer matrix parameters. The same is valid relative
equations (39, 40) in the paper \cite{20}.  }

Averaging over $\psi$ leads to equation (5) with parameters
$$
D=2\epsilon^2 \left\langle\cos^2{(\psi\!-\!\gamma)}
\right\rangle\,,
\quad
$$
$$
\gamma_1 D=\epsilon^2 \left\langle 1\!-
\! 2 \cos^2{(\psi\!-\!\gamma)}\right\rangle
\,,\quad
\eqno(34)
$$
$$
\gamma_2 D=
\Delta\left\langle \vphantom{R^2}
\cos{(\psi\!-\!\beta)}\right \rangle
-\epsilon^2 \left\langle \vphantom{R^2}
\sin{(\psi\!-\!\gamma)}\right\rangle\,,
$$
while for the 'drift velocity' in (6) one has
$$
v=2 \Delta\left\langle \vphantom{R^2}
\cos{(\psi\!-\!\beta)}\right \rangle +
$$
$$
+2 \epsilon^2
\left\langle \sin^2{(\psi\!-\!\gamma)}
-\sin{(\psi\!-\!\gamma)}\right\rangle\,.
\eqno(35)
$$
The averaging in Eqs.(34,\,\,35) is produced over the stationary
distribution $P(\psi)$.

\begin{center}
{\bf 4. Dependence on the properties of the ideal leads}
\end{center}

Let the Fermi momentum $k$ in the ideal leads is different from
the Fermi momentum $\bar k$ in the disordered system, while
the boundary between them is abrupt. Then the edge matrices in
the forbidden band can be chosen in the form
$$
T_l =  \left ( \begin{array}{cc} l\,\, & l^* \\
l^* & l\,\, \end{array} \right)\,,\qquad
T_r= \left ( \begin{array}{cc} r\,\, & r^* \\
r^* & r\,\, \end{array} \right)\,,
\eqno(36)
$$
$$
l=\frac{1}{2}\left(1+\frac{\kappa}{ik} \right)\,,\qquad
r=\frac{1}{2}\left(1+\frac{ik}{\kappa} \right)\,,
$$
leading to the results (18) for matrices $T_{\epsilon_n}$
and $T_{\delta}$ with parameters
$$
\alpha= -\frac{1}{2}
\left(\frac{k}{\kappa}-\frac{\kappa}{k} \right)\delta=
\frac{\kappa^2-k^2}{2k}\,a_0 \,,
$$
$$
\beta= \frac{\pi}{2} \,,\quad \gamma=-\frac{\pi}{2}, \quad
\epsilon^2=\bar \epsilon^2 \left(\frac{\kappa}{k}\right)^2=
\frac{W^2}{4 k^2 a_0^2}\, \,,
\eqno(37)
$$
$$
\Delta= \frac{1}{2}
\left(\frac{k}{\kappa}+\frac{\kappa}{k} \right)\delta=
\frac{\kappa^2+k^2}{2k}\,a_0 \,,
$$
which are the regular functions of the energy  ${\cal E}=-\kappa^2$
and can be analytically continued to the allowed band. Then
for parameters $D$ and $v$ one has
$$
D=2\epsilon^2 \langle\sin^2{\psi}\rangle\,,
\eqno(38)
$$
$$
v=2 \Delta\langle \sin{\psi} \rangle
+2 \epsilon^2 \langle  1\!-\! \cos{\psi}\rangle
-2 \epsilon^2 \langle\sin^2{\psi}\rangle\,,
$$
while the equation for the stationary distribution $P(\psi)$
accepts the form
$$
\epsilon^2 \left(1\!-\!\cos{\psi}\right)^2 P'_\psi+
\, \epsilon^2 \sin{\psi}
(1\!-\! \cos{\psi})  P+
$$
$$
+ \left(\alpha\! -\! \Delta\cos{\psi} \,
\right) P =C_0 \,.
\eqno(39)
$$
The change of variables in Eq.39
$$
\omega=-{\rm ctg}\,\psi/2
\eqno(40)
$$
and renormalization of probability $P\to P(1\!+\!\omega^2)/2$,
following from $P(\psi)d\psi=P(\omega)d\omega$, reduce it to the
simple form
$$
P'_\omega+P(b+a\omega^2)=C_0 \,,
\eqno(41)
$$
where
$$
a=\frac{\alpha-\Delta}{2\epsilon^2}\,,\qquad
b=\frac{\alpha+\Delta}{2\epsilon^2}\,
\eqno(42)
$$
or inversely
$$
\alpha=\epsilon^2(b+a)\,,\qquad
\Delta=\epsilon^2(b-a)\,.
\eqno(43)
$$
Equation (41) can be integrated in quadratures, but
this quadrature is practically useless. It is more
effective to investigate the transformation
properties. If $P_{a,b}(\omega)$ is a solution of Eq.41, then
the following relation is valid
$$
P_{a,b}(\omega)=s^{-1} P_{as^3,bs}(\omega/s)\,.
\eqno(44)
$$
It can be established, making the change $\omega=s\tilde \omega$ and
reducing the obtained equation to the initial form by
redefinition of parameters $\tilde a =a s^{3}$, $\tilde b
=b s$; then $P_{a,b}(\omega)$ coincides with
$P_{\tilde a,\tilde b}(\tilde \omega)$ to the constant factor, which
is established from normalization.
Using the relation
$$
ab=\frac{\alpha^2-\Delta^2}{4\epsilon^4}=
-\frac{\delta^2}{4\epsilon^4}\,,
\eqno(45)
$$
one can see that the scale transformation $a \to a s^3$,
$b \to b s$ leads to renormalization
$\epsilon\to \tilde \epsilon$, where
$$
\tilde \epsilon= \epsilon \, s^{-1}  =
\bar \epsilon \,\frac{\kappa}{k} s^{-1}\,.
\eqno(46)
$$
Substitution of (37) to (43) gives the initial values of the
parameters $a$ and $b$
$$
a=-\frac{k}{\kappa}\frac{\delta}{2\epsilon^2} \,,\qquad
b=\frac{\kappa}{k}\frac{\delta}{2\epsilon^2}   \,,
\eqno(47)
$$
while the relations (33) allow to establish
the change of parameters $\alpha\to \tilde
\alpha$, $\Delta\to \tilde \Delta$  in the result
of the scale transformation
$$
\tilde \alpha=\frac{1}{2}\left( \frac{\kappa}{k} s^{-1} -
 \frac{k}{\kappa} s \right) \delta\,,\quad
\tilde \Delta=\frac{1}{2}\left( \frac{\kappa}{k} s^{-1} +
\frac{k}{\kappa} s \right) \delta \,.
\eqno(48)
$$
Relations (46) and (48) show that transformation of
all parameters $\alpha$, $\Delta$, $\epsilon^2$ entering the
evolution equations reduces to the change
$$
\frac{k}{\kappa} \to
\frac{k}{\kappa} s \,,
\eqno(49)
$$

\noindent
which is equivalent to renormalization of the Fermi
momentum in the ideal leads. Inversely, variation of
the properties of the ideal leads results in the scale
transformati- on of the distribution $P(\omega)$.

\begin{center}
{\bf 5. Influence of the delta  potential
on interfaces} \end{center}

If there is the delta potential on interfaces between the
disordered system and the ideal leads, then the edge matrices
(36) transform to the form
$$
T_l =  \left ( \begin{array}{cc}\!\!\! l & \!\! l_1^*\!\!\! \\
\! l^* & l_1 \!\!\end{array} \right),\quad
l=\frac{ik\!+\!\kappa\!-\!\kappa_1}{2ik},\,\,
l_1=\frac{ik\!+\!\kappa\!+\!\kappa_1}{2ik},
\eqno(50)
$$
$$
T_r= \left ( \begin{array}{cc}\!\! r\,\, & \!\! r^* \!\!\!\\
\!\! r_1^* &\!\! r_1 \!\!\end{array} \right)\,,\quad
r=\frac{ik\!+\!\kappa\!-\!\kappa_2}{2\kappa}, \,\,
r_1=\frac{ik\!+\!\kappa\!+\!\kappa_2}{2\kappa},
$$
where $\kappa_1$ and $\kappa_2$ corresponds to the jumps of
the logarithmic derivative of the wave function on the left
and right interfaces. The condition  $T_r T_l=1$ is realized
for  $\kappa_2=-\kappa_1$, and can be accepted without the
loss of generality (see Footnote 3). Then the matrix
$T_{\epsilon_n}$ remains unchangeable, while $T_{\delta}$
accept the form  (18) with parameters
$$
\alpha= -\frac{k^2-\kappa^2+\kappa_1^2}{2\kappa k} \delta
 \,,
$$
$$
\Delta \cos{\beta}=-
\frac{\kappa_1}{\kappa } \delta  \,,
\eqno(51)
$$
$$
\Delta \sin{\beta}=
\frac{k^2+\kappa^2-\kappa_1^2}{2\kappa k} \delta
 \,,
$$
which, as previously, are the regular functions of the
energy ${\cal E}=-\kappa^2$ (due to $\delta=\kappa a_0$).
Thereby, we again have $\gamma=-\pi/2$, while the
parameter $\beta$ becomes different from the value $\pi/2$. The
change of variables (40) transforms equation (33) to the
form
$$
P'_\omega+P(b+c\omega+a\omega^2)=C_0 \,,
\eqno(52)
$$
with parameters
$$
a=\frac{\alpha\!-\!\Delta\sin{\beta}}{2\epsilon^2},\,\,
b=\frac{\alpha\!+\!\Delta\sin{\beta}}{2\epsilon^2},\,\,
c=-\!\frac{\Delta\cos{\beta}}{\epsilon^2},
\eqno(53)
$$
which accept the following form after substitution of the
physical values (51):
$$
a=-\frac{k}{\kappa}\frac{\delta}{2\epsilon^2}, \,\,\,
b=\frac{\kappa^2-\kappa_1^2}{\kappa k}
\frac{\delta}{2\epsilon^2},\,\,\,
c=\frac{\kappa_1}{\kappa}\frac{\delta}{2\epsilon^2}.
\eqno(54)
$$
A solution of Eq.\,52 satifies to relation
$$
P_{a,b,c}(\omega)=s^{-1} P_{as^3,bs,cs^2}(\omega/s)\,,
\eqno(55)
$$
and the scale transformation of $P(\omega)$ as previously
corresponds to a change of the Fermi momentum $k$ in the ideal
leads.

It is easy to see, that the shift $\omega\to \omega\!-\!\omega_0$
of the variable $\omega$  allows to reduce Eq.\,52 to
the form (41), which is expressed by the relation
$$
P_{a,b,c}(\omega)=P_{\tilde a,\tilde b,0}(\omega\!+\!\omega_0)\,,
\eqno(56)
$$
$$
\omega_0=- \frac{\kappa_1}{k}\,,\quad
\tilde a=a=-\frac{k}{\kappa}\frac{\delta}{2\epsilon^2} \,,\quad
\tilde b=\frac{\kappa}{k}\frac{\delta}{2\epsilon^2}
  \,,
$$
where the values of parameters $\tilde a$ and $\tilde b$
corresponds to the situation with $\kappa_1=0$. As a result,
the appearance of the delta potential on interfaces leads
to translation of the distribution $P(\omega)$ to the
quantity $\omega_0$, proportional to the amplitude
of the delta potential.

\begin{figure*}
\centerline{\includegraphics[width=5.6 in]{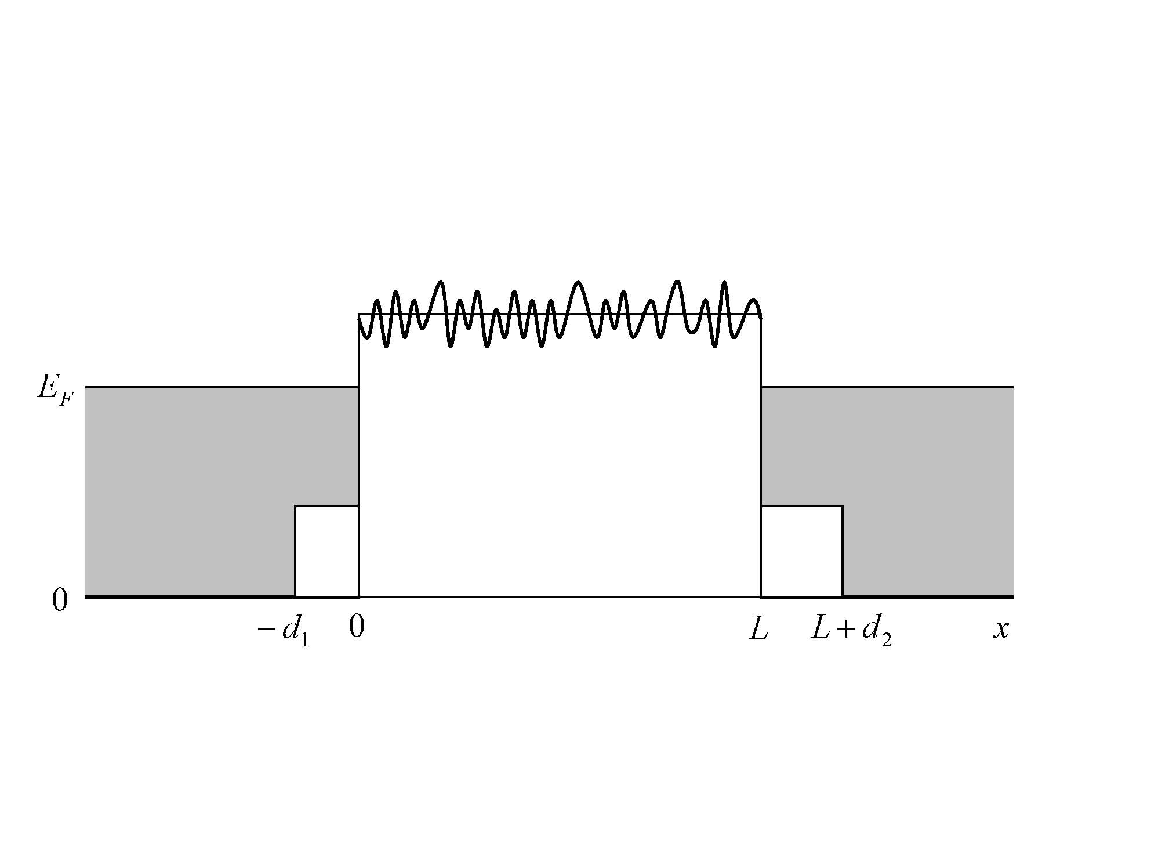}} \caption{
The model to describe the smearing of interfaces.
}\label{fig4}
\end{figure*}
\begin{center}
{\bf 6. Smearing of interfaces }
\end{center}

For energies in the forbidden band, the disordered system
in the absence of impurities reduces to the potential barrier.
The smearing of interfaces can be simulated by incorporating
the layers of a metal with the Fermi momentum $k_1$
and thickness $d_1$ and $d_2$ on the boundaries of the system
(Fig.4). The edge matrices $T_l$ and $T_r$ for such model
are given in Appendix 2. The condition $T_r T_l=1$
is satisfied,
if the following constraints are imposed on $d_1$ and $d_2$
$$
k_1(d_1\!+\!d_2)=2\pi m\,,\quad k(d_1\!+\!d_2)=2\pi n \,,
\eqno(57)
$$
where $m$ and $n$ are integers. The matrix $T_\delta$
is the transfer matrix of the general form, and its calculation
is not very actual. For the matrix $T_{\epsilon_n}$
one obtains the expression (18) with parameters
$$
K=\frac{\kappa{\cal R}}{k_1}\,, \quad
\gamma={\rm arcsin} \left(
\frac{\sin{\alpha}}{\cal R} \right) -\frac{k}{k_1}\alpha
-\frac{\pi}{2}\,,
\eqno(58)
$$
where
$$
{\cal R}={\cal P}\!+\!{\cal Q}\cos{\alpha} \,,\quad
\alpha=k_1(d_2-d_1)\,,
$$
$$ {\cal P}=\frac{k_1^2\!+\!k^2}{2k k_1}\,, \quad
{\cal Q}=\frac{k_1^2\!-\!k^2}{2k k_1}\,.
\eqno(59)
$$
The value of $\gamma$, different from $-\pi/2$, is obtained
for $d_1\ne d_2$. Since the parameter $\gamma$ enters both
in expressions (34,\,35) for $D$ and $v$, and in Eq.\,33
for the stationary distribution $P(\psi)$, it can be
excluded from equations by the shift $\psi\to\psi+\psi_0$,
which reduces it to the value $-\pi/2$; only the
corresponding redefinition of $\beta$ is necessary.
After it, the change of variables (40) leads to equation (52) of
the previous section.

If the Fermi momentum $k_1$ is chosen to be proportional to
$k$, then the parameters ${\cal P}$ and ${\cal Q}$ are
independent of $k$, and the scale transformation of $P(\omega)$
as previously corresponds to renormalization (49) of
the Fermi momentum $k$. In general, the proportionality
$k_1\propto k$ is not realized, but the change of the
properties of the ideal leads corresponds
to the scale transformation of $P(\omega)$ with the more
complicated relation of the momentum $k$ with the scale
factor $s$.

\begin{center}
{\bf 7. General approach to the edge matrices}
\end{center}

Let discuss the general approach to analysis of the
role of the boundary conditions, not related with
the model assumptions. We try to find out the degree
of arbitrariness, permissible in the edge matrices.

As was indicated in Sec.2, the condition $T_r T_l=1$ can be
accepted without the loss of generality. According to (16,17),
we are interested in expressions containing the equal number
of matrices $T_r$ and $T_l$, and so the common factors can
be easily
extracted from one matrix and include into another. Hence, without
the loss of generality the matrices $T_r$ and $T_l$ can be
chosen in the form
$$
T_l =  \left ( \begin{array}{cc}\! a &  b \!\\
\! c & d \! \end{array} \right), \quad
T_r= \left ( \begin{array}{cc}\! d &\!\!\!  -b \! \\
\!\!\! -c \! & a \!\end{array} \right),
\quad  ad-bc=1,
\eqno(60)
$$
accepting the unit value for their determinants. The role
of the edge matrices is the most essential in the forbidden
band: they allow to transform the pseudo-transfer matrix $t$
to the true transfer matrix $T=T_l\, t \,T_r$. The condition
for a matrix $T$ to be the true transfer matrix is
expressed by the following relations between its matrix
elements $T_{ij}$ \cite{21}\,\footnote{\,We give these
relations in the form applicable to quasi-1D systems, when
the elements $T_{ij}$ are in fact matrices.}
$$
T_{11} T_{11}^+ - T_{12} T_{12}^+ =1\,,
$$
$$
T_{22} T_{22}^+ - T_{21} T_{21}^+ =1\,,
\eqno(61)
$$
$$
T_{11} T_{21}^+ - T_{12} T_{22}^+ =0\,,
$$
which follows from the flux conservation.

Let demand that matrices $T_{\delta}$ and $T_{\epsilon}$,
entering (16,17) were true transfer matrices. Then (see
Appendix 3) the admissible  edge matrices form the free
parameter family
$$
\!\! T_l = \! \left ( \begin{array}{cc}\!\!\!
|r_1| {\rm e}^{i\alpha_1}\!  &
\!\!  |r_2| {\rm e}^{i\alpha_2} \!\!\! \!\\
\!\!\!\pm i |r_1| {\rm e}^{-i\alpha_1} \!\!
&\!\! \pm i |r_2| {\rm e}^{-i\alpha_2} \!\!\!\!
\end{array} \right)\! , \,\,
1\!=\! 2 |r_1 r_2| \sin\!{(\alpha_2\!\!-\!\!\alpha_1\!)},
\eqno(62)
$$
where the upper sign is chosen for
$\sin\!{(\alpha_2\!-\!\alpha_1)}\!>\! 0$, and the lower
sign in the opposite case. Then for the matrices
$T_{\delta}$ and $T_{\epsilon}$ we have the expressions
(18) with parameters
$$
\alpha =- 2 \, \delta
|r_1| |r_2| \cos\!{(\alpha_2\!-\!\alpha_1\!)}\,,
$$
$$
\beta=\alpha_1\!+\!\alpha_2\,,
$$
$$
\gamma =2\, \alpha_1 -2\, {\rm arcsin} \,
\frac{|r_2| \cos\!{(\alpha_2\!-\!\alpha_1\!)}}{r} \,,
$$
$$
\Delta = 2 \,\delta |r_1| |r_2| \,,
\eqno(63)
$$
$$
K=r^2 = |r_1|^2 \!+\! |r_2|^2 \!
- 2  |r_1| |r_2| \cos\!{(\alpha_2\!-\!\alpha_1\!)}\,.
$$
Analogous research in the allowed band leads to the
physically evident result: the edge matrices are the
true transfer matrices of  the general form
$$
T_l =  \left ( \begin{array}{cc}\!\!\!
|r_1| {\rm e}^{i\alpha_1}\! &
\!|r_2| {\rm e}^{i\alpha_2} \!\!\! \\
\!\!\! |r_2| {\rm e}^{-i\alpha_2}
\!\! &\!  |r_1| {\rm e}^{-i\alpha_1} \!\!\!
\end{array} \right),\,\,
|r_1|^2- |r_2|^2=1 .
\eqno(64)
$$
In this case, we have the following parameters in the
expressions (18) for $T_{\delta}$ and $T_{\epsilon}$
$$
\alpha =- \delta \left(|r_1|^2+ |r_2|^2 \right) \,,
$$
$$
\beta=\pi/2+\alpha_1\!+\!\alpha_2\,,
$$
$$
\gamma = -\frac{\pi}{2} +
2\, \alpha_1 -2\, {\rm arcsin} \,
\frac{|r_2| \cos\!{(\alpha_1\!-\!\alpha_2\!)}}{r} \,,
$$
$$
\Delta = 2 \,\delta |r_1| |r_2| \,,
\eqno(65)
$$
$$
K=r^2 = |r_1|^2 \!+\! |r_2|^2 \!
- 2  |r_1| |r_2| \cos\!{(\alpha_2\!-\!\alpha_1\!)}\,.
$$
It should be clear from the preceding sections, that the
whole free parameter arbitrariness of the edge matrices can be
realized physically, while the evolution equations derived
in Sec.3 have the most general form.

\begin{figure*}
\centerline{\includegraphics[width=5.6 in]{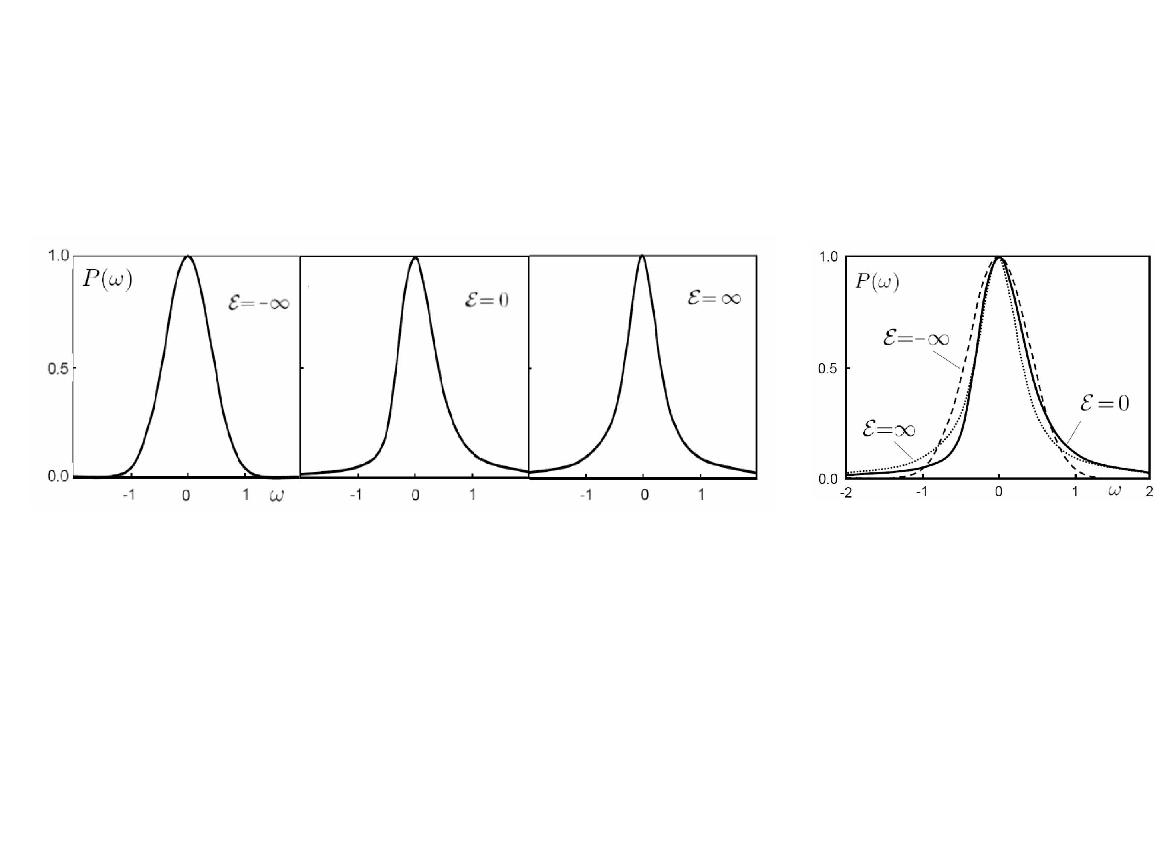}}
\caption{
Distributions $P(\omega)$ for different energies differ
only slightly.
} \label{fig5}
\end{figure*}

\begin{center}
{\bf 8. Distribution $P(\omega)$}
\end{center}

As clear from preceding, the most general results for the
parameters of the evolution equation (5) and the limiting
distribution (6) are given by relations (34,\,\,35), where
averaging over $\psi$ is carried out over the
stationary distribution $P(\psi)$, determined by
equation (33). The shift $\psi\to \psi+\psi_0$ allows to
reduce the parameter $\gamma$ to the value $-\pi/2$, and
after it the change of variables $\omega=-{\rm ctg}\psi/2$
leads to equation (52). Then translation
$\omega\to\omega+\omega_0$ allows to transfer (52) to equation
(41), which by the scale transformation
$\omega\to s\omega$ can be reduced to one of the canonical
form, either with $|a|=|b|$, or with $a=-1$, depending
on the single parameter; this one-parameter freedom is
associated with the dependence on the reduced energy
$\tilde{\cal E}$. Thereby, the form of the distribution
$P(\omega)$ with $\omega=-{\rm ctg}\,\psi/2$ is
determined by the internal properties of the system,
while the change of the boundary conditions transfers it
to the distribution $P(\tilde\omega)$, where the variable
$$
\tilde\omega = \omega_0 - s\, {\rm ctg} \frac{\psi+\psi_0}{2}
\eqno(66)
$$
is related with $\omega$ by the homographic
transformation
$$
\tilde\omega = \frac{A\omega + B}{C \omega +D}\,.
\eqno(67)
$$
According to \cite{15}, the situation \mbox{$|a|\!=\!|b|$}
is distinguished by the fact, that in the deep of the
allowed and forbidden bands parameters of the evolution
equation (5) remain constant till very small length
scales\,\footnote{\,For sufficiently large $L$ these parameters
are always constant, since they are determined by the
stationary distribution $P(\psi)$.}. The case \mbox{$a\!=\!b$},
which is realized in the allowed band, corresponds to the
'natural' ideal leads (Sec.1). Under condition
\mbox{$-a\!=\!-b\gg 1$}  (in the deep of the allowed band)
one can omit the term $P'_\omega$ in Eq.\,41, and obtain
$$
P(\omega) =
\frac{1}{\pi}\,\frac{1}{1+\omega^2}\,,
\eqno(68)
$$
which correspond to $P(\psi)={\rm const}$, i.e. the random phase
approximation.  For $-a\!=\!b\gg 1$ (in the deep of the
forbidden band) one can neglect the constant $C_0$ in Eq.\,41
near the maximum of distribution and, setting
$\omega=-1+\tilde\omega$, obtain $P'_\omega\approx
-2|a|\tilde\omega P$, which gives the Gaussian distribution
$$
P(\omega) = \sqrt{ \frac{|a|}{\pi} }\,
\exp\left\{-|a|\tilde\omega^2 \right\}\,,
\eqno(69)
$$
localized near \mbox{$\omega\!=\!-1$}; it corresponds to the
Gaussian distribution $P(\psi)$, localized near $\pi/2$ \cite{15}.
If the condition $|a|=|b|$ is violated in the allowed band,
it leads to appearance of the oscillations of distribution
$P(\rho,\psi)$, related with the fact that the matrix $T$
in the absence of impurities reduces to the transfer matrix of
the potential barrier, which becomes transparent, if its width
$L$ corresponds to the semi-integer number of the de Broglie
waves (analogously to blooming in optics) \cite{15}. If the
condition $|a|=|b|$ is violated in the deep of the forbidden
band, then the distribution $P(\rho,\psi)$ is
localized near the value of $\psi$, different from $\pi/2$,
which leads to its slow relaxation to the latter value.

Near the edge of the initial band, it is convenient to
reduce equation (41) to the canonical form with
\mbox{$a=-1$}. Its solution with $b=0$ corresponds to the
'critical' region for the smoothed Anderson transition
\cite{15} and can be presented in the form (see Appendix 4)
$$
P(\omega) = A_0 \int\limits_0^\infty dt {\rm e}^{-t} t^{-1/3}
\frac{\omega}{\omega^3+3t}\,,
\eqno(70)
$$
where $A_0$ is determined from normalization.  For large
$\omega$ we have $P(\omega)\sim \omega^{-2}$, which is the
general property  (see Appendix 4) and corresponds to the finite
values of $P(\psi)$ on the boundaries of the interval $(0,2\pi)$.

\begin{figure*}
\centerline{\includegraphics[width=7.2 in]{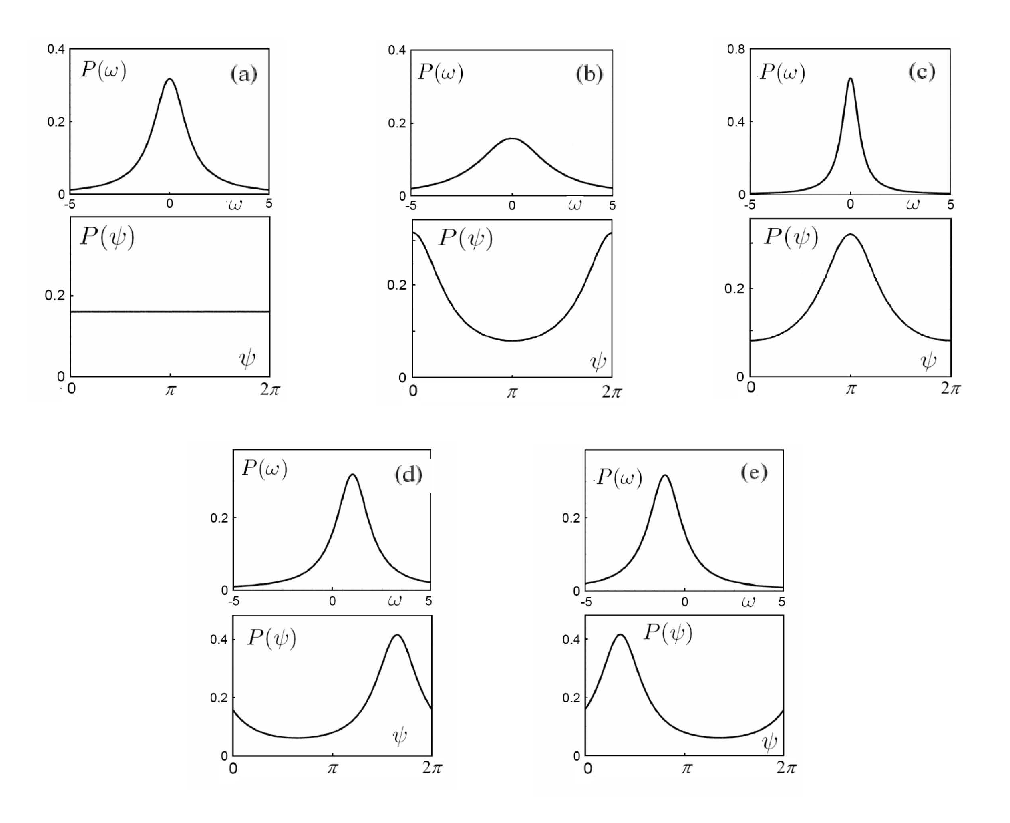}} \caption{
The change of the distribution $P(\psi)$ under the scale
transformation and translation of the function $P(\omega)$,
if the form of the latter corresponds to the random
phase approximation.
} \label{fig6}
\end{figure*}

If by the scale transformation and translation we provide
the maximum of the distribution $P(\omega)$ at $\omega=0$
with the unit value in it, then distributions (68--70)
differ not very essentially (Fig.5), witnessing on
the weak dependence of $P(\omega)$ on the reduced energy
$\tilde {\cal E}$.

\begin{center}
{\bf 9. External phase distribution}
\end{center}

If the distribution $P(\omega)$ is known, then its
change under variation of the boundary condition reduces
to the scale transformation and translations. The corresponding
change of the external phase distribution $P(\psi)$ is easily
predictable on the qualitative level and illustrated in Fig.6.
The distribution $P(\psi)$ is uniform (Fig.6,a), if $P(\omega)$
has a form (68) and corresponds to the random phase
approximation.
Widening of the distribution $P(\omega)$ leads to
localization of $P(\psi)$  near the edges of the interval
$(0,2\pi)$ (Fig.6,b), while narrowing leads to
localization of $P(\psi)$ in the middle of the interval
$(0,2\pi)$.  (Fig.6,c).
A shift of $P(\omega)$ to right or to left leads to
appearance of the maximum of $P(\psi)$ in the right or left
part of the interval $(0,2\pi)$ (Fig.6,d,e). If the parameter
$\gamma$ is different from $-\pi/2$, it leads to translation
$\psi\to\psi+\psi_0$ and the presented dependences are valid not
in the interval $(0,2\pi)$, but in the interval
$(\psi_0,2\pi+\psi_0)$.

\begin{figure*}
\centerline{\includegraphics[width=5.6 in]{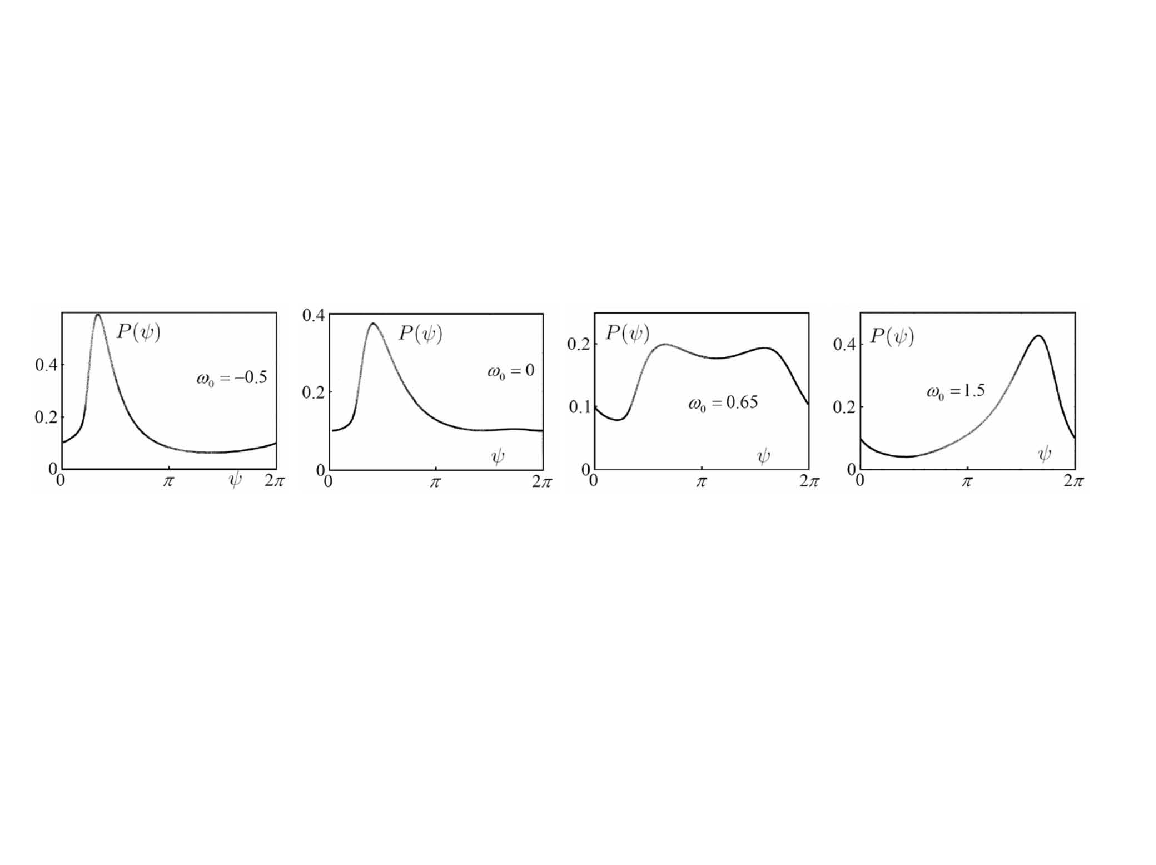}}
\caption{
The change of the distribution $P(\psi)$ under translations
of the critical distribution  $P(\omega)$.
} \label{fig7}
\end{figure*}

If the distribution $P(\omega)$ is accepted not in the
form (68), but in the Gaussian form (69), then Fig.6
does not change qualitatively, but $P(\psi)$ will not be
constant in Fig.6,a. The asymmetrical form of the function
$P(\omega)$ for ${\cal E}=0$, corresponding to the
critical distribution (Fig.5), leads to a new
qualitative effect: its translations
$\omega\to\omega+\omega_0$ leads to appearance of the
two-humped distribution $P(\psi)$  in a certain
interval of the $\omega_0$ values (Fig.7).

\begin{center}
{\bf 10. Internal phase distribution}
\end{center}

Let return to the case \mbox{$\beta=\pi/2$},
\mbox{$\gamma=-\pi/2$}, considered in Sec.4, and
produce the change of variables $\omega=-{\rm ctg}\,\psi/2$
in the complete evolution equation (31). If the typical
values of $\rho$ are sufficiently large, then the
evolution equation for $P(\rho,\omega)$ accepts the form
$$
\frac{\partial P}{2\partial L}=
\left\{\vphantom{\frac{1}{2}}
\epsilon^2  \frac{4\omega^2}
{\left(\vphantom{\omega^2} 1+\omega^2\right)^2} \rho^2 P'_\rho
-\Delta  \frac{2\omega}{1+\omega^2} \rho P +
\right.
$$
$$ \left.
+\epsilon^2 \frac{2} {1+\omega^2} \rho P  +\ldots
\right\}'_\rho +
\epsilon^2\,\left\{ \vphantom{\frac{1}{2}}  P'_\omega
    + \left(b+a\omega^2\right) P\,
 \right\}'_\omega \,.
\eqno(71)
$$
Making the change $\omega=s\tilde\omega$ and redefinition of
parameters $\tilde a=as^3$, $\tilde
b=bs$, $\tilde\epsilon^2 =\epsilon^2 s^{-2}$, we have
$$
\frac{\partial P}{2\partial L}=
\left\{\vphantom{\frac{1}{2}}
\tilde\epsilon^2 s^2 \frac{4s^2\tilde\omega^2}
{\left(\vphantom{\omega^2} 1+s^2\tilde\omega^2\right)^2}
\rho^2 P'_\rho
-\tilde\Delta  \frac{2s\tilde\omega}
{1+s^2\tilde\omega^2} \rho P +
\right.
$$
$$ \left.
+\tilde\epsilon^2 s^2 \frac{2}
{1+s^2\tilde\omega^2} \rho P  +\ldots
\right\}'_\rho +
\tilde\epsilon^2\,\left\{ \vphantom{\frac{1}{2}}
P'_{\tilde\omega}
+ \left(\tilde b+\tilde a\tilde\omega^2\right) P\,
 \right\}'_{\tilde\omega} \,.
\eqno(72)
$$
The latter term in Eq.\,72 has the same form as in Eq.\,71,
but the other terms are  not invariant. According to Eq.5,
the parameter $D$ is determined by the coefficient
before $\rho^2 P'_\rho$, and equations (71,\,\,72) lead
to two expressions
$$
D=2\epsilon^2 \left\langle
\frac{4\omega^2}
{\left(\vphantom{\omega^2} 1+\omega^2\right)^2}
\right\rangle_{a,b}  \,,
\eqno(73)
$$
$$
D=2\tilde\epsilon^2 s^2 \left\langle
\frac{4s^2\omega^2}
{\left(\vphantom{\omega^2} 1+s^2\omega^2\right)^2}
\right\rangle_{\tilde a,\tilde b}   \,,
\eqno(74)
$$
which are equivalent due to relation (44). If Eq.74 is
accepted as a definition of the diffusion coefficient $D$,
then its invariance in respect to the change of the boundary
conditions is fulfilled automatically\,\footnote{\,Such definition 
was implicitly accepted in the preliminary variant of this paper 
[Phil. Mag. Lett. {\bf 102}, 255 (1922)] and used in 
transformation of Eq.53 to Eq.54. Necessity of agreement of two 
definitions (see Eq.75 below) was not realized in this 
communication.}. However, the change
of $a$, $b$, $\epsilon$ to $\tilde a$, $\tilde b$,
$\tilde\epsilon$ in Eq.\,71 leads to the analogous changes in
Eq.\,73, which together with Eq.\,74 gives the relation
$$
\left\langle
\frac{4\omega^2}
{\left(\vphantom{\omega^2} 1+\omega^2\right)^2}
\right\rangle_{\tilde a,\tilde b} =
 s^2 \left\langle
\frac{4s^2\omega^2}
{\left(\vphantom{\omega^2} 1+s^2\omega^2\right)^2}
\right\rangle_{\tilde a,\tilde b}  \,,
\eqno(75)
$$
which is not fulfilled for arbitrary $s$, and valid only
for a certan 'correct' choice of the scale factor.
It easy to understand that this 'correct' choice
corresponds to the 'internal' phase distribution.
Analogous considerations can be given in respect to
the parameter $v$ in the distribution (6), which is
determined by the coefficient before $\rho P$.

\begin{figure}
\centerline{\includegraphics[width=3.4 in]{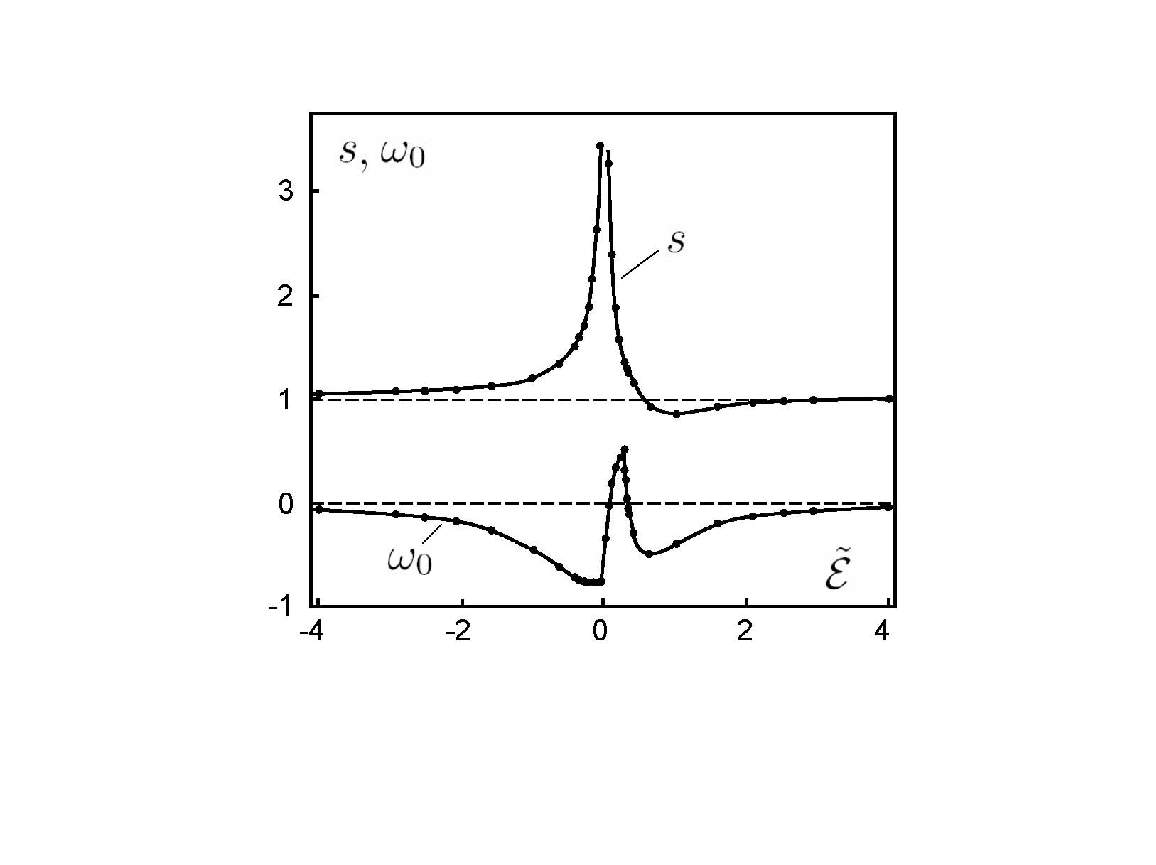}} \caption{
Behavior of the scale factors $s$ and the translational shift
 $\omega_0$ as functions of the reduced energy $\tilde{\cal E}$.
} \label{fig8}
\end{figure}
\begin{figure}
\centerline{\includegraphics[width=3.4 in]{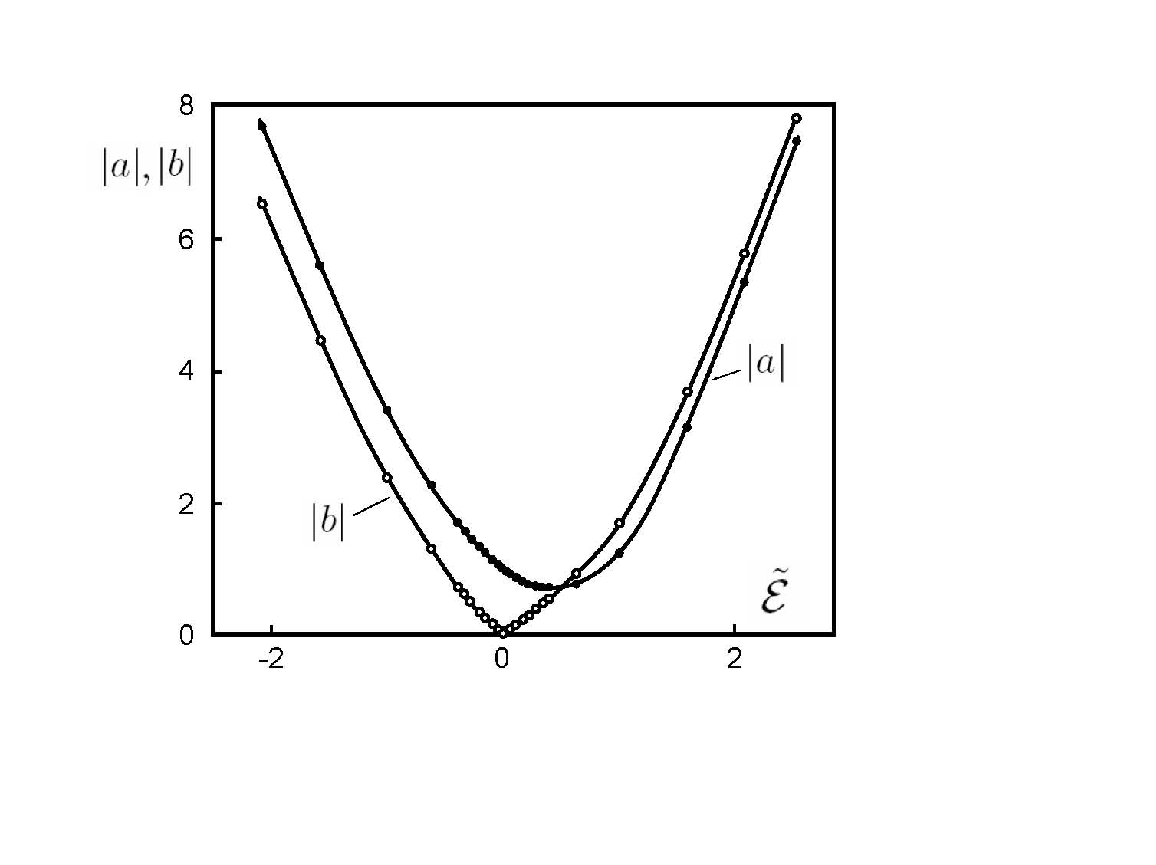}}
\caption{ Parameters $a$ and $b$ for the internal phase
distribution as functions of the reduced energy
$\tilde{\cal E}$.  }
\label{fig9}
\end{figure}

Let the initial values $a_0$, $b_0$ of the parameters $a$, $b$
are chosen from the condition $|a_0|=|b_0|$, which is 'natural'
beyon the vicinity of the initial band edge \cite{15}. If
$P(\omega)$ is the solution of equation (41) with  parameters
$a_0$, $b_0$, then replacement $\omega\to
s(\tilde\omega+\omega_0)$ allows to provide the correct values of
$D$ and $v$, following from analysis of moments, by the proper
 choice of $s$ and $\omega_0$. The obtained distribution
 $P(\tilde\omega)$ after return to the variable $\psi$ will
determine the internal phase distribution. Behavior of $s$ and
$\omega_0$ as a function of the reduced energy $\tilde{\cal E}$
is shown in Fig.8. The scale factor  $s$ is diverging at
\mbox{$\tilde{\cal E}\to 0$}, and the renormalized parameters
$a=a_0 s^3$ and $b=b_0 s$ become not be bounded by the relation
\mbox{$|a|=|b|$} (Fig.9), which is realized only for large
$|\tilde{\cal E}|$, i.e. in the deep of the allowed and
forbidden bands. The physical meaning of this fact is more
explicit from Fig.10, demonstrating the dependence of the
Fermi energy $\tilde{\cal E}_{ext}$ in the ideal leads against
the Fermi energy $\tilde{\cal E}$ in the disordered system.
The minimum of $\tilde{\cal E}_{ext}$ is reached at the
value $\tilde{\cal E}=\tilde{\cal E}_0=0.28$, which can be
interpreted as a renormalized band edge, shifted due to
fluctuations of a random potential. This conclusion is confirmed
by the fact, that for \mbox{$\tilde{\cal E}<\tilde{\cal E}_0$}
the translational shift $\omega_0$ moves to the complex
plane, which has a simple physical sense. For $\tilde{\cal
E}=\tilde{\cal E}_0$ we have
\mbox{$\tilde{\cal E}_{ext}=\tilde{\cal E}_0$} (Fig.10)
and the shifted  band edge is below the Fermi level in the ideal
leads for $\tilde{\cal E}>\tilde{\cal E}_0$ and upper it
for $\tilde{\cal E}<\tilde{\cal E}_0$. In the latter case
the internal properties of the system are described not by
true, but the pseudo-transfer matrix, and the distribution
$P(\psi)$ becomes complex-valued. The detailed study of the latter
is not very actual, and the 'internal' distribution $P(\psi)$,
presented below  for $\tilde{\cal E}<\tilde{\cal E}_0$,
is in fact the external distribution, which is closest to
internal; it practically  coincides with the modulus of the
complex distribution $P(\psi)$. The latter follows from the fact,
that the shift of the parameter $\omega_0$ to the complex plane
is comparatively small and allows analytical investigation (see
Appendix 5): it is illustrated by Fig.11, which shows the best
fit of the parameter $v$ (with exact fitting of $D$),
possible for real values of $\omega_0$.

\begin{figure}
\centerline{\includegraphics[width=3.4 in]{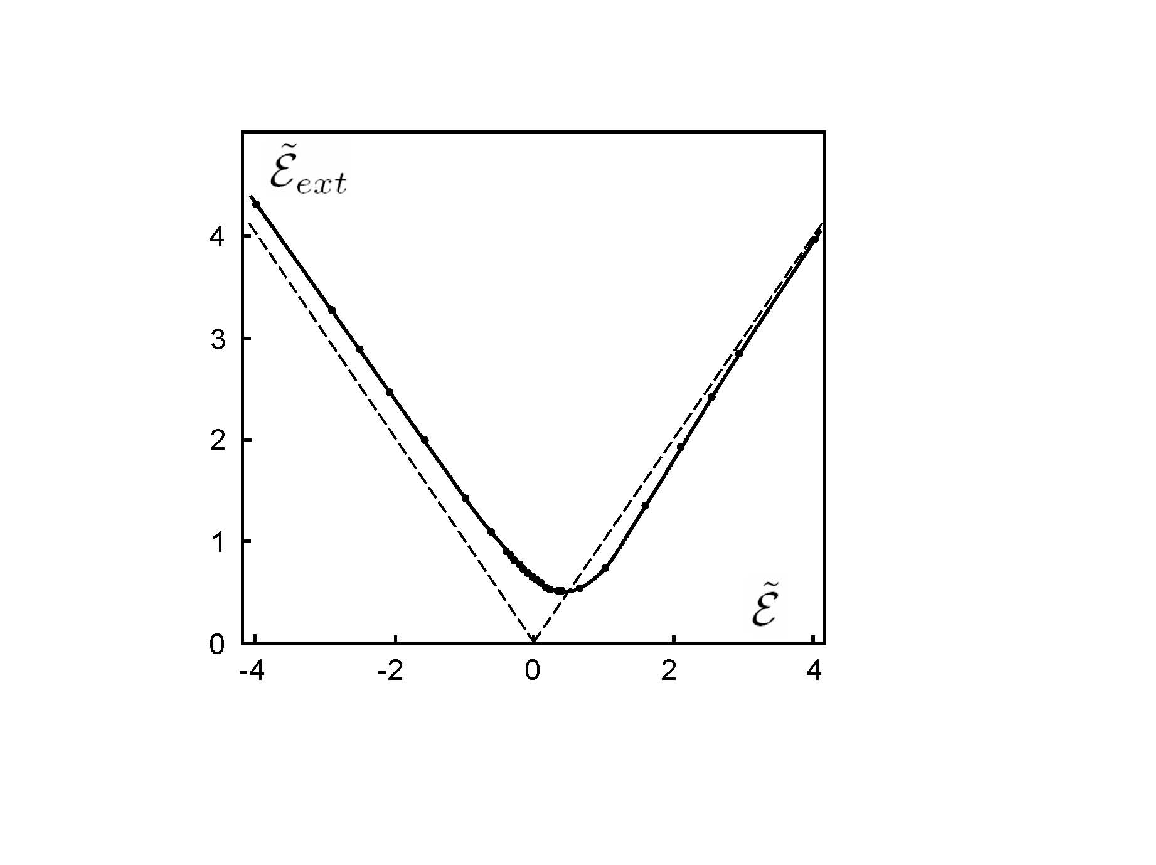}}
\caption{
Dependence of the Fermi energy $\tilde{\cal E}_{ext}$ in the
ideal leads on the Fermi energy in the disordered system
 $\tilde{\cal E}$ for the internal phase distribution.
 The dashed line corresponds to the dependence
 $\tilde{\cal E}_{ext}=|\tilde{\cal E}|$.
} \label{fig10}
\end{figure}
\begin{figure}
\centerline{\includegraphics[width=3.4 in]{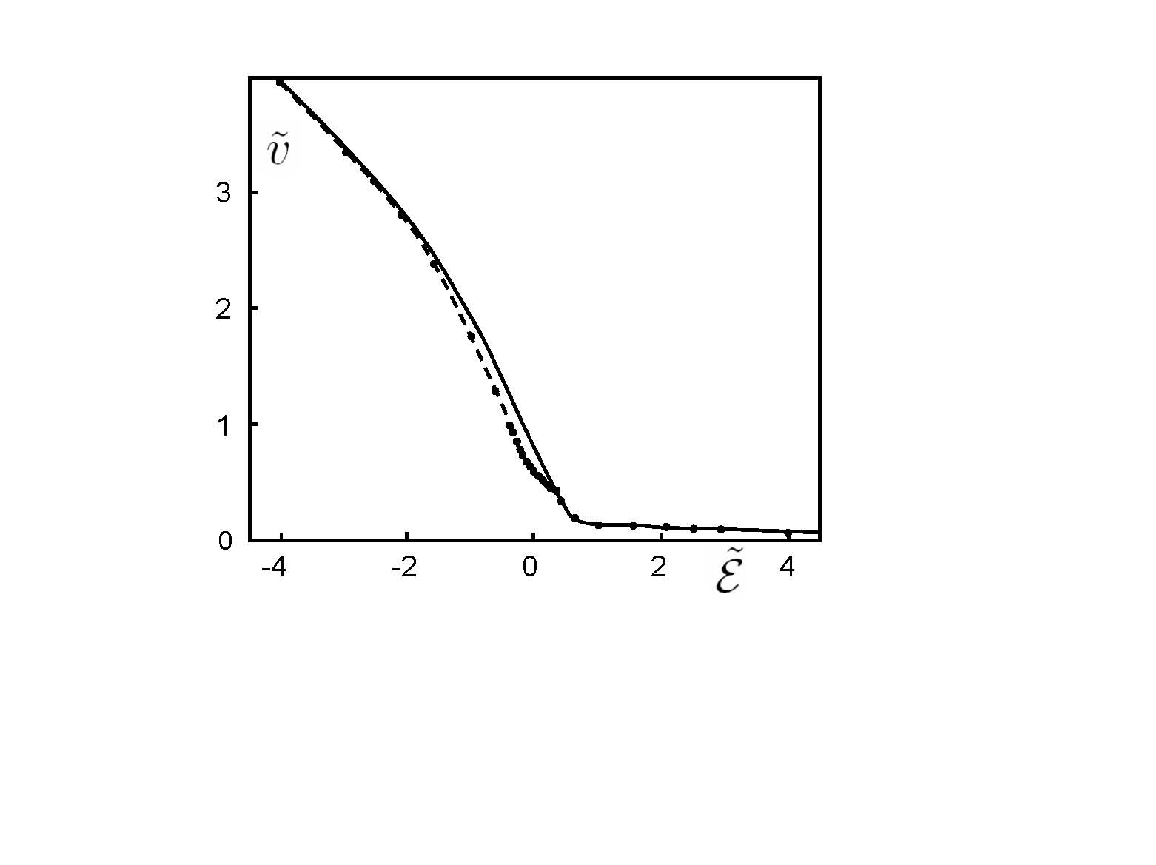}} \caption{
The best fit of the parameter $v$ for the real values of the
translational shift  $\omega_0$.
} \label{fig11}
\end{figure}

Evolution of the internal distribution $P(\psi)$ (with a given
reservation) under variation of the reduced energy $\tilde{\cal E}$
is shown in Fig.12. The first row of figures corresponds to
variation of $\tilde{\cal E}$ from large positive values till
the value $\tilde{\cal E}=0.63$, corresponding to the right
minimum of $\omega_0$ in Fig.8; here the distribution
$P(\omega)$ is close to the Lorentz form (68), while the change
of $P(\psi)$ is mainly related with translation of $P(\omega)$
to the left direction (Fig.6,e).
The third row of Fig.12  corresponds to
variation of $\tilde{\cal E}$ from large negative values till
the value $\tilde{\cal E}=-0.10$, corresponding to the left
minimum of $\omega_0$ in Fig.8; here the distribution
$P(\psi)$ is localized near $\pi/2$ and gradually spreads
with growth of $\tilde{\cal E}$, while $P(\omega)$
changes from the Gaussian form (69) to the critical form (70).
The second row of Fig.12  corresponds to
variation of $\tilde{\cal E}$ between two minima
of $\omega_0$ in Fig.8; here the distribution $P(\omega)$
is close to the critical form (70), while the change
of $P(\psi)$ is mainly related with translational shift of
$P(\omega)$ and corresponds to Fig.7. The fourth and
ninth dependences of Fig.12 are close to each other, since
the form of $P(\omega)$ is close to critical, while the
translational shifts $\omega_0$ are approximately equal.

Let discuss the mechanism of realization of the internal
phase distribution. As was discussed in papers \cite{15,22,23},
the distribution $P(\rho)$ of the Landauer resistance undergoes
aperiodic oscillations, related with the fact that the form of
$P(\rho)$ depends essentially on the several first moments,
while the moments $\left\langle\rho^n\right\rangle$ are
oscilating functions of $L$: the $n$th moment is
determined by a superposition of $n$ discrete
harmonics. The relative amplitude of oscillations decays
exponentially on the scale $L\sim \xi$, and they become
inessential for large $L$. The analogous situation takes place
for the complete distribution $P(\rho,\omega)$;\,\footnote{\,The
distribution $P(\rho)$ is determined by the even moments of the
transfer matrix elements $T_{ij}$. The odd moments should be
taken into account in the analysis of $P(\rho,\omega)$.}
\begin{figure*}
\centerline{\includegraphics[width=7.2 in]{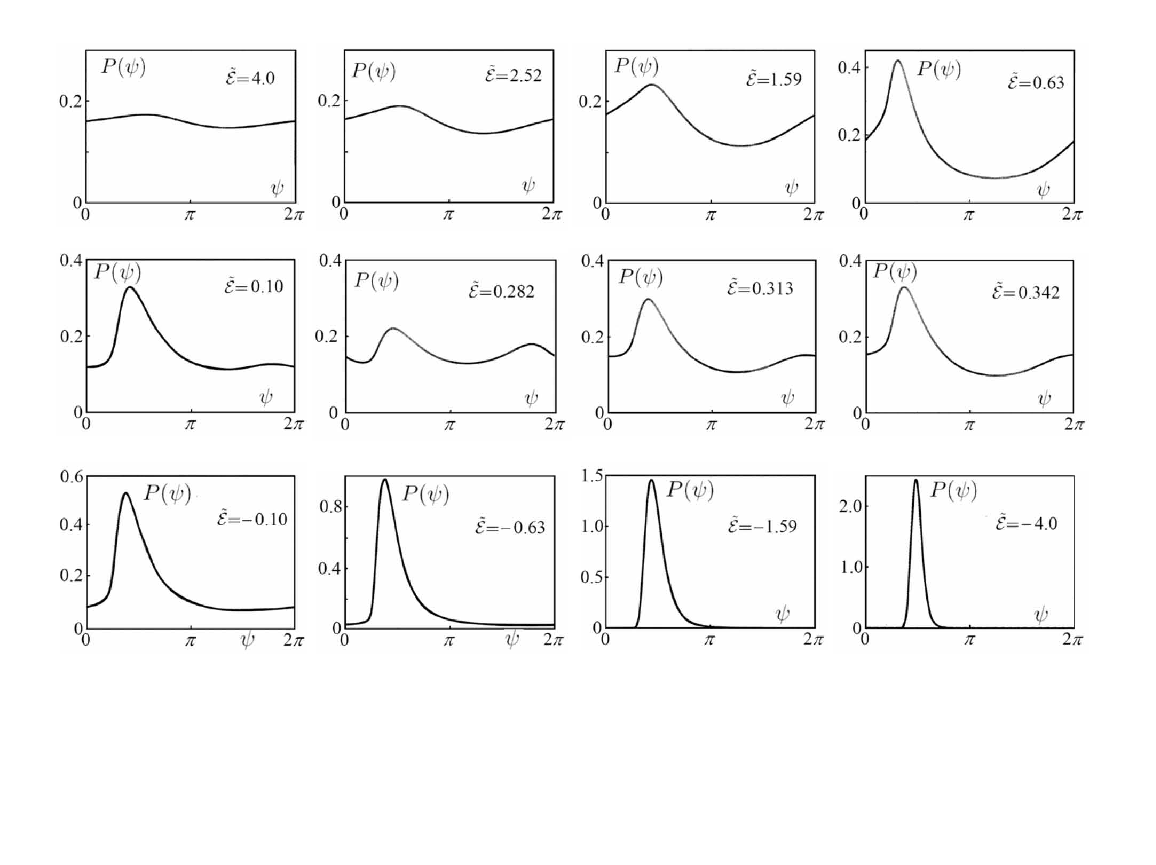}}
\caption{Evolution of the internal phase distribution
under variation of the reduced energy $\tilde{\cal E}$.
} \label{fig12}
\end{figure*}
%
correspondingly, both the mean of $\omega$ and its dispersion
are oscillating. The latter are directly related with the
translational shift $\omega_0$ and the scale factor $s$, which
become oscillating on the scale $L\alt \xi$, tending to
'correct' constant values at large $L$: these 'correct' values
correspond to the internal phase distribution.

\begin{center}
{\bf 11. Conclusion}
\end{center}

In the present paper we have derived the evolution equation for
the mutual distribution $P(\rho,\psi)$ of the Landauer resistance
$\rho$ and the phase variable $\psi$. For large $L$, this
equation allows the separation of variables, which provides the
existence of the stationary distribution $P(\psi)$, determinating
the coefficients in the evolution equation (5) for $P(\rho)$.

In the result of the present analysis we come to a very simple
picture. The phase $\psi$ appears to be a "bad" variable, while
the "correct" variable is $\omega=-{\rm ctg}\,\psi/2$. The
form of the stationary distribution $P(\omega)$ is determined by
the internal properties of the system and does not depend on the
boundary conditions.
Variation of the boundary conditions leads to three
effects:  (a) the scale transformation $\omega\to s\omega$,
which is mainly related with the change of properties
of the ideal leads, attached to the system; (b)
translation $\omega \to \omega+\omega_0$, which is mainly
determined by existence of the delta-potential on the
boundaries of the system; (c) translation $\psi\to\psi+\psi_0$,
related with smearing of interfaces.
The boundary conditions essentially affect the 'external' phase
distribution, which enters the evolution equations, but do not
affect the limiting distribution $P(\rho)$ in the region of
large $L$, which allows to say on existence of the hidden
symmetry in the evolution equations.
The limiting distribution $P(\rho)$ has the log-normal form,
which can be established under the most general
conditions\,\footnote{\,More precisely, we consider a random
potential of small amplitude with short-ranged correlations
for energies near the edge of the initial band.
If the energy is comparable with the bandwidth, there
arise the effects of commensurability of the Fermi mimentum
$\bar k$ with the lattice constant $a_0$
\cite{100,101}, which can result in complication of the
situation \cite{102,103}.}. The limiting distribution $P(\rho)$
is determined by the 'internal' distribution $P(\psi)$, which is
obtained from the stationary distribution $P(\omega)$ with the
proper choice of the scale factor $s$ and translational shifts
$\omega_0$ and $\psi_0$.

The discussed problems are not restricted by 1D
systems, and analogous difficulties arise in the
studies of the Lyapunov exponents in the
framework of the generalized version \cite{16} of
the Dorokhov--Mello--Pereyra-Kumar equation \cite{17,18}. The
minimal Lyapunov exponent determinates the critical properties of
the Anderson transition (it is clear from the well-known
numerical algorithm, see references in  \cite{16}), and the
analogous hidden symmetry can be essential in the studies of the
latter.

\begin{center}
{\it Appendix 1.} {\bf To separation of variables in the equation
(31).}
\end{center}

For separation of variables in the eigenvalue problem
$$
\hat L \, P(\rho,\psi)=\lambda \, P(\rho,\psi)
\eqno(A.1)
$$
the operator  $\hat L$ should be represented as a sum of
two operators $\hat L_\rho +\hat M_\psi$, depending only on
$\rho$ and $\psi$ correspondingly.

Conditions for separation of variables in the equation (31)
appear to be essentially weaker. In the limit of large $L$,
when the typical values of $\rho$ are sufficiently large,
one can set $R=2$ to have the following structure for
equation (31)
$$
\frac{\partial P}{\partial L}=
\left\{\vphantom{L^2} \hat L_{\rho,\psi} P \right\}'_\rho +
\left\{ \vphantom{L^2} \hat M_{\psi} P \right\}'_\psi \,.
\eqno(A.2)
$$
Setting $P=P(\rho) P(\psi)$ and dividing by $P(\rho)$, one has
$$
-\frac{\partial P(\psi)}{\partial L} +
\left\{ \vphantom{L^2} \hat M_{\psi} P(\psi) \right\}'_\psi
= \qquad
\eqno(A.3)
$$
$$ \qquad
=\frac{P(\psi)}{P(\rho)} \frac{\partial P(\rho)}{\partial L}
-\frac{1}{P(\rho)}
\left\{\vphantom{L^2} \hat L_{\rho,\psi} P \right\}'_\rho
\,.
$$
The left-hand side is independent of $\rho$, and can be
considered as a certain function $F(\psi)$. Then
$$
P(\psi) \frac{\partial P(\rho)}{\partial L}
-\left\{\vphantom{L^2} \hat L_{\rho,\psi} P \right\}'_\rho
= F(\psi) P(\rho)
\eqno(A.4)
$$
and integration over $\rho$ gives $F(\psi)\equiv 0$,
since the left-hand side turns to zero, while the integral over
$P(\rho)$ is equal to unity due to normalization. As a result,
the left-hand side and right-hand side of Eq.\,$(A.3)$ turn to
zero independently, and the equation for $P(\psi)$ is splitted
off.  Correspondinly, the averages of function of $\psi$,
entering (34,\,\,35), are determined by the stationary
distribution $P(\psi)$, satisfying to equation (33).

\begin{center}
{\it Appendix 2.} {\bf Edge matrices for the smeared
interfaces}
\end{center}

The edge matrices for the model represented in Fig.4 can
be chosen in the form
$$
T_l= \left ( \begin{array}{cc}
{\rm e}^{ik d_1} & 0 \\ 0 & {\rm e}^{-ik d_1}
\end{array} \right)\,T_a\,
\left ( \begin{array}{cc}
{\rm e}^{-ik_1 d_1} & 0 \\ 0 & {\rm e}^{ik_1 d_1}
\end{array} \right)\,T_c
\eqno(A.5)
$$
$$
T_r= T_{\tilde c} \left ( \begin{array}{cc}
{\rm e}^{-ik_1 d_2} & 0 \\ 0 & {\rm e}^{ik_1 d_2}
\end{array} \right)\,T_{\tilde a}\,
\left ( \begin{array}{cc}
{\rm e}^{ik d_2} & 0 \\ 0 & {\rm e}^{-ik d_2}
\end{array} \right)\,,
$$
where
$$
T_a= \left ( \begin{array}{cc}
a_1 & a_2 \\ a_2 & a_1 \end{array}
\right)\,,\qquad  a_1=\frac{k+k_1}{2k}\,,\quad
a_2=\frac{k-k_1}{2k}\,,
$$
$$
T_c= \left ( \begin{array}{cc}
c & c^* \\ c^* & c \end{array}
\right)\,,\qquad  c=\frac{ik_1+\kappa}{2ik_1}\,,
\eqno(A.6)
$$
$$
T_{\tilde a}= \left ( \begin{array}{cc}
\tilde a_1 & \tilde a_2 \\ \tilde a_2 & \tilde a_1 \end{array}
\right)\,,\qquad  \tilde a_1=\frac{k_1+k}{2k_1}\,,\quad
\tilde a_2=\frac{k_1-k}{2k_1}\,,
$$
$$
T_{\tilde c}= \left ( \begin{array}{cc}
\tilde c & \tilde c^* \\ \tilde c^* & \tilde c \end{array}
\right)\,,\qquad  \tilde c=\frac{\kappa+ik_1}{2\kappa}\,,\quad
$$
Comparison with \cite{15} shows that $T_{a}$ and $T_{\tilde a}$
are the edge matrices for the boundary between two metals, while
$T_c$ and $T_{\tilde c}$ are the edge matrices for the
boundary between a metal and a dielectric.

Composing the product $T_l T_r$, we find that relations
$T_{c}T_{\tilde c}=1$ and $T_{a}T_{\tilde a}=1$ leads
to essential simplifications, and under condition (57)
this product reduces to the unit matrix.

For calculation of $T_\epsilon$ we can write
$$
T_\epsilon= \left ( \begin{array}{cc}
1 & 0 \\ 0 & 1 \end{array} \right)\,+\,\epsilon \,
T_l \left ( \begin{array}{cc}
\! 1 & \! 1 \! \\ \!\!\! -1 & \!\!\! -1\! \end{array} \right) T_r
\eqno(A.7)
$$
and use the relations
$$
T_c \left ( \begin{array}{cc}
1 & 1 \\ -1 & -1 \end{array} \right) T_{\tilde c} =
\frac{\kappa}{ik_1} \left ( \begin{array}{cc}
 1 \! & 1 \! \\ \!-1 & \!-1 \end{array} \right) \,,
$$
$$
\left ( \begin{array}{cc}
\! {\rm e}^{-ik_1 d_1} \!\!\! & 0
\\ \! 0 \! & \!\! {\rm e}^{ik_1 d_1} \!\!
\end{array} \right)\,
\left ( \begin{array}{cc}
1 & 1 \\ -1 & -1 \end{array} \right)
\left ( \begin{array}{cc}
\! {\rm e}^{-ik_1 d_2} \!\!\! & 0
\\ 0 & \!\! {\rm e}^{ik_1 d_2} \!\!
\end{array} \right)\,=
$$
$$ 
\qquad \qquad \qquad\qquad 
=\left ( \begin{array}{cc}
1 & {\rm e}^{i\alpha} \\ -{\rm e}^{-i\alpha} & -1
\end{array} \right)
\eqno(A.8)
$$
$$
T_a\left ( \begin{array}{cc}
\!\! 1 \! & \! {\rm e}^{i\alpha} \!
\\ \!\! -{\rm e}^{-i\alpha} \! & \! -1 \!
\end{array} \right) T_{\tilde a}=
\qquad\qquad\qquad\qquad
$$
$$=
\left ( \begin{array}{cc} \!\! {\cal P}+{\cal Q}\cos{\alpha} \!
& \!\! {\cal Q}+{\cal P}\cos{\alpha}+i\sin{\alpha} \!\! \\
\!\! -{\cal Q}-{\cal P}\cos{\alpha} +i\sin{\alpha} \! & \!\!
-{\cal P}-{\cal Q}\cos{\alpha} \!\!  \end{array} \right) \,,
$$
which with the use of equality  ${\cal P}^2\!=\!{\cal Q}^2+1$
allow to reduce $T_\epsilon$ to the form  (18).

\begin{center}
{\it Appendix 3.} {\bf Degree of arbitrariness for the
edge matrices} \end{center}

Let accept the form (60) for the edge matrices $T_l$ and $T_r$,
and demand that the matrix $T_\delta=T_l\, t_\delta \,T_r$
was the true transfer matrix, satisfying conditions (61);
it is sufficient to impose the latter for small $\delta$
and obtain the relations
$$
ad+bc+a^*d^*+b^* c^*=0\,,\qquad ab+c^* d^*=0\,.
\eqno(A.9)
$$
The first relation together with the unit condition for
the determinant leads to more simple relations
$$
2\,{\rm Re}\, ad=1\,, \qquad 2\,{\rm Re}\, bc=-1\,.
\eqno(A.10)
$$
Setting
$$
a=|a| {\rm e}^{i\alpha},\quad
b=|b| {\rm e}^{i\beta},\quad
c=|c| {\rm e}^{i\gamma},\quad
d=|d| {\rm e}^{i\delta}\,,
\eqno(A.11)
$$
we have the complete set of conditions
$$
2 |b| |c| \cos{(\beta\!+\!\gamma)}=-1\,,
$$
$$
2 |a| |d| \cos{(\alpha\!+\!\delta)}=1\,,
$$
$$
|a| |b| = |c| |d|\,,
\eqno(A.12)
$$
$$
{\rm e}^{i\alpha+i\beta+i\gamma+i\delta}=-1\,,
$$
$$
|a| |d| \sin{(\alpha\!+\!\delta)} -|b| |c| \sin{(\beta\!+\!\gamma)}=0\,.
$$
It is easy to see  that
$\sin{(\alpha\!+\!\delta)}=\sin{(\beta\!+\!\gamma)}$,
$\cos{(\alpha\!+\!\delta)}=-\cos{(\beta\!+\!\gamma)}$;
then from the last relation $(A.12)$ we have $|a| |d| = |b| |c|$,
which gives $|a|=|c|$, $|b| = |d|$ together with the third
relation $(A.12)$. As a result, the condition for $T_\delta$
to be the true transfer matrix determines
4-parameter family for the matrix $T_l$,
$$
T_l =  \left ( \begin{array}{cc}\!\!\! |a| {\rm e}^{i\alpha}\! & \!
|b| {\rm e}^{i\beta} \!\!\! \\ \!\!\! |a| {\rm
e}^{i\gamma} \!\! &\! |b| {\rm e}^{i\delta} \!\!\!
\end{array} \right), \,\,
$$
$$
{\rm e}^{i\alpha+i\beta+i\gamma+i\delta}=-1\,, \quad
1=-\! 2 |a| |b| \cos\!{(\beta\!+\!\gamma\!)}.
\eqno(A.13)
$$
Calculating the matrix $T_\epsilon$
$$
T_\epsilon =  \left ( \begin{array}{cc}
1-\bar\epsilon (a\!-\!b)(c\!-\!d)  &
\bar\epsilon (a\!-\!b)^2 \\ -\bar\epsilon (c\!-\!d)^2
&1+\bar\epsilon (a\!-\!b)(c\!-\!d)  \end{array} \right), \,\,
\eqno(A.14)
$$
and demanding that it was the true transfer matrix, we obtain
$$
(a\!-\!b)(c\!-\!d)+(a^*\!-\!b^*)(c^*\!-\!d^*)=0\,,
$$
$$
(a\!-\!b)^2+(c^*\!-\!d^*)^2=0\,,
\eqno(A.15)
$$
Introducing the parametrization
$$
(\!a-\!b)=r {\rm e}^{i\theta/2}\,,\quad
(c\!-\!d)=r_1 {\rm e}^{i\varphi/2}\,,
\eqno(A.16)
$$
we have  $r=r_1$, $\varphi=\pi-\theta +2\pi n$, so
$$
T_\epsilon =  \left ( \begin{array}{cc}
1-i \bar\epsilon r^2 (-1)^n &
\bar\epsilon r^2 {\rm e}^{i\theta} \\
\bar\epsilon r^2 {\rm e}^{-i\theta} &
1+i \bar\epsilon r^2 (-1)^n
\end{array} \right), \,\,
\eqno(A.17)
$$
where $n$ is integer. Taking into account relations
$$
r^2=|a\!-\!b|^2 = |a|^2 \!+\! |b|^2 \!
- 2  |a| |b| \cos\!{(\alpha\!-\!\beta\!)}\,,
$$
$$
r^2=|c\!-\!d|^2 = |a|^2 \!+\! |b|^2 \!
- 2  |a| |b| \cos\!{(\gamma\!-\!\delta\!)}\,,
\eqno(A.18)
$$
we have $(\alpha\!-\!\beta\!)=\pm (\gamma\!-\!\delta\!)+2\pi k$.
The upper sign leads to contradiction, since
$\beta\!\!+\!\gamma\!=\pi/2+\pi k$
and the left-hand side of $(A.12)$ turns to zero.
For the lower sign we have
$\gamma\!=\pi/2-\alpha+\pi k$, $\delta\!=\pi/2-\beta+\pi k'$,
where $k$ and $k'$ should be chosen even for
$\sin\!{(\!\beta\!\!-\!\alpha\!)}>0$ and odd in the opposite
case. As a result, we come to representation (62) for the matrix
$T_l$ with 3-parameter arbitrariness.  For even $k$, $k'$
we have even $n$ in the expression $(A.17)$ for $T_\epsilon$,
and it reduces to (18) with parameters $K$ and  $\gamma$,
given by Eq.63.  For odd $k$, $k'$ we have odd $n$ in $(A.17)$,
and to reduce it to the form (18) we need to make the inessential
change of sign for $\bar\epsilon$.

\begin{center}
{\it Appendix 4.} {\bf Distribution $P(\omega)$ for
${\cal E}=0$} \end{center}

Differentiating (41) and suggesting the power-law asymptotics
$P(\omega)\sim \omega^{-\alpha}$ for \mbox{$\omega\to\infty$},
we obtain \mbox{$\alpha=2$}. Considering
corrections to the asymptotics in the form of a series
over inverse powers of $\omega$,
$$
P(\omega) = \omega^{-2} \sum\limits_{n=0}^\infty A_n
\omega^{-n}\,,
\eqno(A.19)
$$
we have the reccurence relations
$$
A_0=1\,,\quad A_1=0\,,\quad A_2=-b/a\,,
$$
$$
a A_{n+1}=-b A_{n-1}+n A_{n-2}\,,\quad n\ge 2\,.
\eqno(A.20)
$$
For the critical distribution (${\cal E}=0$) we have  $b=0$,
and the coefficients of the series can be found explicitly
$$
P(\omega) = \frac{1}{\omega^2} \sum\limits_{k=0}^\infty
\frac{\Gamma(k+2/3)}{\Gamma(2/3)}
\left( \frac{3}{a\omega^3}  \right)^k  \,.
\eqno(A.21)
$$
The divergent series can be summed in the Borel sense
\cite{24}, if one write the gamma function $\Gamma(k+2/3)$
in the form of the defining integral and summing the
arising geometrical progression. Setting  $a\!=\!-1$
and including the normalization factor, we obtain (70).

Expressions $(A.19)$, $(A.20)$ are useful for numerical
integration of equation (41) for arbitrary $b$. For
$|\omega|>5$ the function $P(\omega)$ is well approximated
by the series $(A.19)$, interpreted in the asymptotical
sense and summed till the minimal term. Such approximation
can be used as an initial condition at large positive
$\omega$ for integration in the direction of diminishing of
$\omega$.\,\footnote{\,Attempts to integrate in the direction
of increasing of $\omega$ meet instabilities, relating with
existence of the growing exponent, being a solution of
equation (41) for $C_0=0$.} If the
standard procedures with the accuracy control \cite{25} are used,
the solution automatically comes to the correct asymptotics at
$\omega\to-\infty$.

\begin{center}
{\it Appendix 5.} {\bf Behavior of $s$ and $\omega_0$ in
the forbidden band}
\end{center}

In the deep of the forbidden band, the constant $C_0$ in (41)
is exponentially small, and the distribution $P(\omega)$ can
be accepted in the form
$$
P(\omega) = {\rm const} \exp\left\{-b\omega-a\omega^3/3
  \right\} \,.
\eqno(A.22)
$$
In the case $-a=b\gg 1$ the distribution has a
maximum at $\omega=-1$, in whose vicinity it reduces to the
Gaussian form (69). Calculation of averages in (38) gives
the results
$$
\left\langle \sin{\psi} \right\rangle=
\left\langle -\frac{2\omega}{1+\omega^2} \right\rangle
\approx
\left\langle 1-\frac{\tilde\omega^2}{2} \right\rangle=
 1-\frac{1}{4|a|}  \,,
$$
$$
\left\langle \sin^2{\psi} \right\rangle=
\left\langle \frac{4\omega^2}{(1+\omega^2)^2} \right\rangle
\approx
\left\langle 1-\tilde\omega^2 \right\rangle\approx 1\,,
$$
$$
\left\langle \cos{\psi} \right\rangle=
\left\langle \frac{\omega^2-1}{1+\omega^2} \right\rangle
\approx
\left\langle -\tilde\omega \right\rangle\approx 0\,,
\eqno(A.23)
$$
and for parameters $D$ and $v$ one has
$$
D=2\bar\epsilon^2\,,\quad
v=2\delta-\bar\epsilon^2\,,
\eqno(A.24)
$$
which agree with results from the analysis of
moments \cite{15} within accuracy; here $\tilde\omega$ is
deviation from the maximum, and the fact is used that
$\epsilon$ reduces to $\bar\epsilon$ for
$|a|=|b|$.

To reproduce $D$ and $v$ with higher accuracy, let make the scale
transformation and translation of the distribution $(A.22)$
with $-a=b$; then for $s=1-\tilde s$,
$\omega_0=\tilde s+\tilde\omega_0$ and $\tilde
s,\,\tilde\omega_0\ll 1 $ one has
$$
D=2\bar\epsilon^2
\left(1+2\tilde s-\bar\epsilon^2/\delta \right) \,,
\eqno(A.25)$$
$$
v=2\delta-\bar\epsilon^2 +
\left(\tilde s^2-\tilde\omega_0^2 \right)\delta
-\bar\epsilon^2\left(2\tilde\omega_0+2\tilde s+
9\bar\epsilon^2/4\delta\right)\,,
$$
where the shift of the maximum and small deviations from the
Gaussian form are taken into account. Comparison with
values from the analysis of moments \cite{15}
$$
D=2\bar\epsilon^2 -\frac{15}{2}\frac{\bar\epsilon^2}{\delta}
\,,
\quad
v=2\delta-\bar\epsilon^2 +
\frac{27}{4}\frac{\bar\epsilon^2}{\delta}\,
\eqno(A.26)
$$
leads to the following results,  if the relation
$\delta/\bar\epsilon^2=4|\tilde{\cal E}|^{3/2}$ is
taken into account,
$$
s=1+\frac{11}{32}\frac{1}{|\tilde{\cal E}|^{3/2}}
\,,\quad
\omega_0=\frac{-19\pm i\sqrt{215}}{32|\tilde{\cal E}|^{3/2}}
\,,
\eqno(A.27)
$$
which explains the behavior of curves in the left-hand part
of Fig.8. The translational shift $\omega_0$ appears to be
complex-valued; one can easily verify for the Gaussian
distribution, that neglection of the small imagionary part of
$\omega_0$ is equivalent to taking the modulus of the
complex distribution $P(\omega)$.

\end{document}